\pdfoutput=1
\documentclass[useAMS,usenatbib]{mn2e}
\usepackage{apjfonts}
\usepackage{graphicx}
\usepackage{amsmath}
\usepackage{amssymb}
\usepackage{ctable}
\usepackage{fixltx2e}
\usepackage{hyperref}
\hypersetup{colorlinks=true,linkcolor=blue,citecolor=blue,filecolor=blue,urlcolor=blue}

\newcommand{\be}{\begin{equation}}
\newcommand{\ee}{\end{equation}}
\newcommand{\ba}{\begin{eqnarray}}
\newcommand{\ea}{\end{eqnarray}}

\newcommand{\Ms}{M_{\ast}}
\newcommand{\Msun}{{\rm M}_{\sun}}

\newcommand{\nc}{n_{\rm th}}
\newcommand{\tb}{t_{\rm lookback}}
\newcommand{\z}{|Z|}
\newcommand{\zh}{Z_H}
\newcommand{\cm}{{\rm cm}}
\newcommand{\pc}{{\rm pc}}
\newcommand{\kpc}{{\rm kpc}}
\newcommand{\age}{{\rm age}}
\newcommand{\yr}{{\rm yr}}
\newcommand{\Gyr}{{\rm Gyr}}
\newcommand{\dkpc}{{\rm dex\,kpc^{-1}}}
\newcommand{\Fe}{{\rm Fe}}
\newcommand{\Mg}{{\rm Mg}}
\newcommand{\MM}{{\rm M}}
\newcommand{\HH}{{\rm H}}
\newcommand{\dd}{{\rm d}}
\newcommand{\aFe}{[\alpha/\Fe]}
\newcommand{\MgFe}{[\Mg/\Fe]}
\newcommand{\FeH}{[\Fe/\HH]}
\newcommand{\MH}{[\MM/\HH]}
\newcommand{\rgrad}{\dd[\MM/\HH]/\dd R}
\newcommand{\vgrad}{\dd[\MM/\HH]/\dd\z}
\newcommand{\sech}{{\rm sech}}

\newcommand{\aj}{AJ}
\newcommand{\apj}{ApJ}
\newcommand{\apjl}{ApJ}
\newcommand{\apjs}{ApJS}
\newcommand{\mnras}{MNRAS}
\newcommand{\aap}{A\&A}
\newcommand{\araa}{ARA\&A}
\newcommand{\aapr}{A\&ARv}
\newcommand{\nat}{Nature}

\title[The Stellar Disk of a Simulated MW-Mass Galaxy]
{The Structure and Dynamical Evolution of the Stellar Disk of a Simulated Milky Way-Mass Galaxy}

\author[X. Ma et al.]{
\parbox[t]{1.0\textwidth}{
   Xiangcheng Ma,$^1$\thanks{E-mail: \href{mailto:xchma@caltech.edu}{xchma@caltech.edu}}
   Philip F. Hopkins,$^1$
   Andrew R. Wetzel,$^{1,2,3}$\thanks{Caltech-Carnegie Fellow}
   Evan N. Kirby,$^{4}$ \\
   Daniel Angl{\'e}s-Alc{\'a}zar,$^5$
   Claude-Andr{\'e} Faucher-Gigu{\`e}re,$^5$
   Du{\v s}an Kere{\v s}$^6$ \\ and
   Eliot Quataert$^7$
  }
  \vspace{5pt} \\
  $^1$TAPIR, MC 350-17, California Institute of Technology, Pasadena, CA 91125, USA \\ 
  $^2$Carnegie Observatories, 813 Santa Barbara Street, Pasadena CA 91101, USA \\
  $^3$Department of Physics, University of California, Davis, CA 95616, USA \\
  $^4$Department of Astrophysics, MC 249-17, California Institute of Technology, Pasadena, CA 91125, USA \\
  $^5$Department of Physics and Astronomy and CIERA, Northwestern University, 2145 Sheridan Road, Evanston, IL 60208, USA \\
  $^6$Department of Physics, Center for Astrophysics and Space Sciences, University of California at San Diego, 9500 Gilman Drive, La Jolla, CA 92093, USA \\
  $^7$Department of Astronomy and Theoretical Astrophysics Center, University of California Berkeley, Berkeley, CA 94720, USA
}

\pagerange{\pageref{firstpage}--\pageref{lastpage}}
\date{Draft version \today}

\begin{document}
\maketitle
\label{firstpage}

\begin{abstract}
We study the structure, age and metallicity gradients, and dynamical evolution using a cosmological zoom-in simulation of a Milky Way-mass galaxy from the Feedback in Realistic Environments project. In the simulation, stars older than 6\,Gyr were formed in a chaotic, bursty mode and have the largest vertical scale heights (1.5--2.5\,kpc) by $z=0$, while stars younger than 6\,Gyr were formed in a relatively calm, stable disk. The vertical scale height increases with stellar age at all radii, because (1) stars that formed earlier were thicker `at birth', and (2) stars were kinematically heated to an even thicker distribution after formation. Stars of the same age are thicker in the outer disk than in the inner disk (flaring). These lead to positive vertical age gradients and negative radial age gradients. The radial metallicity gradient is negative at the mid-plane, flattens at larger disk height $\z$, and turns positive above $\z\sim1.5\,\kpc$. The vertical metallicity gradient is negative at all radii, but is steeper at smaller radii. These trends broadly agree with observations in the Milky Way and can be naturally understood from the age gradients. The vertical stellar density profile can be well-described by two components, with scale heights 200--500\,pc and 1--1.5\,kpc, respectively. The thick component is a mix of stars older than 4\,Gyr which formed through a combination of several mechanisms. Our results also demonstrate that it is possible to form a thin disk in cosmological simulations even with strong stellar feedback.
\end{abstract}


\section{Introduction}
\label{sec:intro}
\citet{gilmore.reid.1983:thick.disk} first discovered that the vertical stellar density profile in the solar neighborhood  in the Milky Way (MW) can be described by two exponential components with scale heights $\sim300\,\pc$ and $\sim1450\,\pc$, respectively, and identified them as the thin disk and the thick disk. Such a two-component profile is also seen in external edge-on disk galaxies \citep[e.g.][]{yoachim.dalcan.2006:thick.disk,comeron.2011:thick.disk.ngc.4244,comeron.2012:thick.disk.s4g}. However, it remains unclear whether the thin and thick disks are two distinct components or one single structure that varies continuously above the disk plane.

Several mechanisms have been proposed to explain the formation of the thick disk, despite the ambiguity of whether it is a discrete component or not. Some popular scenarios, all motivated by theory or observations, include: (1) kinematic heating from a pre-existing thin disk during minor mergers \citep[e.g.][]{quinn.1993:merger.heating,kazantzidis.2008:thick.disk.merger,hopkins.2008:disk.heating,villalobos.2008:thick.disk.merger,purcell.2009:disk.heating.merger,qu.2011:thick.disk.merger}, (2) star formation at high redshift from chaotic gas accretion during hierarchical assembly \citep{brook.2004:thick.disk.form} or in a turbulent, gas-rich disk \citep{bournaud.2009:thick.disk.form,haywood.2013:thick.disk.form}, (3) radial migration of stars from the inner disk to the outer disk \citep{schonrich.binney.2009:structure,loebman.2011:thick.migration}, and (4) accretion of stars from SMC-like satellites \citep{abadi.2003:disk.structure}. Nonetheless, it is still unclear which mechanism (or combination of mechanisms) is responsible for the formation of thick disks in the MW and other galaxies.

Thanks to spectroscopic surveys of stars in the MW, such as RAVE \citep{steinmetz.2006:rave.dr1}, SEGUE \citep{yanny.2009:segue.data}, APOGEE \citep{allende-prieto.2008:apogee}, and Gaia-ESO \citep{gilmore.2012:ges.description}, one can now combine three-dimensional position, velocity, and chemical abundance information for large samples \citep[for a recent review, see][]{rix.bovy.2013:mw.disk.review}. Many groups have claimed that there are two distinct sub-populations, named $\alpha$-rich and $\alpha$-poor stars, as revealed by the gap in the $\aFe$--$\MH$ plane ($\MH$ represent total stellar metallicity relative to solar abundance) or the bimodality of the $\aFe$ distribution at fixed $\MH$. These two populations are attributed to the thick and thin disks \citep[e.g.][]{adibekyan.2013:harps.fgk.dwarf,bensby.2014:fg.dwarf,anders.2014:chemo.apogee.dr1,nidever.2014:apogee.red.clump,mikolaitis.2014:ges.idr1.chemo,kordopatis.2015:ges.idr2.alpha}. Nonetheless, some samples show a much less significant gap in the $\aFe$--$\MH$ plane than others \citep[e.g.][]{mikolaitis.2014:ges.idr1.chemo,kordopatis.2015:ges.idr2.alpha}, or no gap at all \citep[e.g.][]{boeche.2014:rave.giant.grad}. Also, in some cases, the bimodality appears in certain $\alpha$ elements but disappears in others \citep[e.g.][]{bensby.2014:fg.dwarf,mikolaitis.2014:ges.idr1.chemo}. Such discrepancies are likely due to large uncertainties in abundance measurements in some studies. The bimodality, if real, implies that the MW may have a hiatus in its star formation history at high redshift \citep[e.g.][]{chiappini.1997:two.infall.model,brook.2012:thin.thick.disk.halo,nidever.2014:apogee.red.clump}. Also, it is not clear whether such feature is common in other galaxies.

Various groups have confirmed a negative radial metallicity gradient with a slope $\rgrad\sim-0.06\,\dkpc$ in MW stars near the disk plane (height $\z<0.5\,\kpc$), with $\rgrad$ gradually flattening above the mid-plane and turning positive at and above $\z>1.5\,\kpc$ \citep[e.g.][]{cheng.2012:segue.stellar.grad,boeche.2013:rave.dwarf.grad,boeche.2014:rave.giant.grad,anders.2014:chemo.apogee.dr1,hayden.2014:chemo.apogee,mikolaitis.2014:ges.idr1.chemo}. A negative vertical metallicity gradient is also found in the MW disk from the mid-plane to $\z\sim3\,\kpc$, but the slope varies dramatically in the literature \citep[e.g.][]{carrell.2012:metal.grad.disk.dwarf,boeche.2014:rave.giant.grad,hayden.2014:chemo.apogee}. \citet{hayden.2014:chemo.apogee} found that the vertical metallicity gradient is steeper at inner Galactocentric radii than at outer radii. 

Nevertheless, using the data at a single epoch alone is not sufficient to identify the mechanism for MW thick disk formation. Cosmological simulations of MW analogs are useful for this problem as they allow one to trace the evolution of the galaxy as well as understand the underlying implications in the observational data. For example, \citet{stinson.2013:magicc.mw.disk} found that older stars tend to have larger scale heights but shorter scale lengths than younger stars in their MW analog simulation, which also supported the observationally motivated conjecture in \citet{bovy.2012:mono.structure} that mono-abundance populations (MAPs; stars with certain $\FeH$ and $\aFe$) are good proxies for single age populations \citep[see also][]{matteucci.1990:mw.bulge.star,fuhrmann.1998:mw.disk.halo.star}. Likewise, many authors have also found a two-component disk structure and similar MAP properties in their simulations \citep[e.g.][]{brook.2012:thin.thick.disk.halo,bird.2013:disk.galaxy.assembly,minchev.2013:chemodynamical,martig.2014:disk.map.structure,minchev.2016:map.age.mw.disk}. Most of these simulations show that the thick disk was formed kinematically hot at high redshift, although it has been debated whether heating is important in disk evolution. For instance, \citet{bird.2013:disk.galaxy.assembly} argued that the thick-disk structure is predominantly determined `at birth', while others suggested that kinematic heating at late times is also significant \citep[e.g.][]{minchev.2013:chemodynamical,martig.2014:disk.age.velocity}.

Additionally, \citet{minchev.2013:chemodynamical} developed a chemo-dynamical model of disk galaxy evolution which reconciled the structure, formation history, and the variation of metallicity gradients in the disk \citep[see also][]{minchev.2014:chemodynamical,minchev.2015:thick.disk.formation}. However, \citet{miranda.2016:thick.disk.metal} pointed out that the metallicity gradients in the disk strongly rely on the treatment of (simplified) feedback in these simulations and only certain recipes produced similar behavior to the MW in their simulations. Therefore, it is important to include realistic models of the interstellar medium (ISM) and stellar feedback to understand the formation of galactic disks.

In this paper, we study a simulation from the Feedback in Realistic Environments project \citep[FIRE;][]{hopkins.2014:fire.galaxy}\footnote{\href{http://fire.northwestern.edu}{http://fire.northwestern.edu}}, which produces a disk galaxy with stellar mass similar to the MW at $z=0$, to study the structure and abundance pattern of stars in the galactic disk. We present the structure and dynamical evolution of the stellar disk, compare the metallicity gradients and their variation with recent observations, and show how the metallicity gradients can be understood from the disk structure. The FIRE project is a suite of cosmological zoom-in simulations with detailed models of the multi-phase ISM, star formation, and stellar feedback taken directly from stellar evolution models and it produces reasonable galaxy properties broadly consistent with observations from $z=0$--6, such as the stellar mass--halo mass relation \citep{hopkins.2014:fire.galaxy,feldmann.2016:fire.quenching.letter}, the Kennicutt--Schmidt law \citep{orr.2017:fire.ks.law}, neutral hydrogen covering fractions around galaxies at both low and high redshift (\citealt{faucher.2015:fire.neutral.hydrogen,faucher.2016:fire.neutral.hydrogen}; Hafen et al. 2016), the stellar mass--metallicity relation \citep{ma.2016:fire.mzr}, mass-loading factors of galactic winds \citep{muratov.2015:fire.mass.loading}, metal budgets and CGM metal content \citep{muratov.2016:fire.metal.loading}, galaxy sizes \citep{elbadry.2016:fire.migration}, and the population of satellite galaxies around MW-mass galaxies \citep{wetzel.2016:high.res.mw.letter}. We briefly summarize the simulation in Section \ref{sec:sim}, present our main results in Section \ref{sec:results}, discuss our findings in Section \ref{sec:discussion}, and conclude in Section \ref{sec:conclusion}.

We adopt a standard flat $\Lambda$CDM cosmology with cosmological parameters $H_0=70.2 {\rm\,km\,s^{-1}\,Mpc^{-1}}$, $\Omega_{\Lambda}=0.728$, $\Omega_{m}=1-\Omega_{\Lambda}=0.272$, $\Omega_b=0.0455$, $\sigma_8=0.807$ and $n=0.961$, broadly consistent with observations \citep[e.g.][]{hinshaw.2013:wmap9.cosmo.param,planck.2014:cosmo.param}. 

\begin{table}
\caption{A list of symbols used in this paper.}
\centering
\begin{center}
\begin{tabular}{cl}
\hline
Symbol & Description \\
\hline
$z$ & Redshift \\
$\tb$ & Lookback time \\
age & Stellar age at $z=0$ \\
$X,\,Y,\,Z$ & Cartesian coordinates \\
$R$ & Galactocentric radius \\
$\z$ & Height from the mid-plane \\
$\MH$ & Total metallicity (relative to solar) \\
$\FeH$ & Fe abundance (relative to solar) \\
$\MgFe$ & Mg to Fe abundance ratio (relative to solar) \\
\hline
\end{tabular}
\end{center}
\label{tbl:symbol}
\end{table}%

\begin{table*}
\caption{Lookback time vs redshift.}
\centering
\begin{center}
\begin{tabular}{lcccccccccc}
\hline
Lookback Time [$\tb$, in Gyr] & 0 & 1 & 2 & 3 & 4 & 5 & 6 & 7 & 8 & 10 \\
\hline
Redshift [$z$] & 0 & 0.076 & 0.162 & 0.258 & 0.369 & 0.497 & 0.649 & 0.834 & 1.068 & 1.812 \\
\hline
\end{tabular}
\end{center}
\label{tbl:tb}
\end{table*}%

\section{Simulation and Methods}
\label{sec:sim}
In this work, we perform a case study using galaxy m12i, a disk galaxy with mass comparable to the Milky Way at $z=0$, from the FIRE project. We pick this simulation because it has been well-studied in previous work \citep[e.g.][]{hopkins.2014:fire.galaxy,muratov.2015:fire.mass.loading,muratov.2016:fire.metal.loading,elbadry.2016:fire.migration,daa.2016:fire.baryon.cycle,ma.2017:fire.metallicity.gradient} and has a morphology that is closest to the MW in our suite. A detailed description of the simulations, numerical recipes, and physics included is presented in \citet[][and references therein]{hopkins.2014:fire.galaxy}. We briefly summarize their main features here. The simulation is run using {\sc gizmo} \citep{hopkins.2015:gizmo.code}, in {\sc p-sph} mode, which adopts a Lagrangian pressure-entropy formulation of the smoothed particle hydrodynamics (SPH) equations that improves the treatment of fluid-mixing instabilities \citep{hopkins.2013:psph.code}. 

The cosmological `zoom-in' initial conditions for m12i are adopted from the AGORA project \citep{kim.2014:agora}. The zoom-in region is about one Mpc in radius at $z=0$. The initial particle masses for baryons and dark matter are $m_{\rm b}=5.7\times10^4\,\Msun$ and $m_{\rm dm}=2.8\times 10^5\,\Msun$, respectively. The minimum force softening lengths for gas and star particles are $\epsilon_{\rm gas}=14\,\pc$ and $\epsilon_{\rm star}=50\,\pc$ (Plummer-equivalent). The force softening lengths for the gas particles are fully adaptive \citep{price.monaghan.2007:adapt.soft}. The most massive halo in the zoom-in region has a halo mass of $M_{\rm halo}=1.4\times10^{12}\,\Msun$ and a stellar mass around $\Ms=6\times10^{10}\,\Msun$ at $z=0$.

In our simulation, gas follows a molecular-atomic-ionized cooling curve from 10--$10^{10}$\,K, including metallicity-dependent fine-structure and molecular cooling at low temperatures and high-temperature metal-line cooling followed species-by-species for 11 separately tracked species \citep[H, He, C, N, O, Ne, Mg, Si, S, Ca, and Fe; see][]{wiersma.2009:metal.cooling}. At each timestep, the ionization states and cooling rates are determined from a compilation of {\sc cloudy} runs, including a uniform but redshift-dependent photo-ionizing background tabulated in \citet{faucher.2009:uvb}, and approximate models of photo-ionizing and photo-electric heating from local sources. Gas self-shielding is accounted for with a local Jeans-length approximation, which is consistent with the radiative transfer calculations in \citet{faucher.2010:lya.cooling}.

We follow the star formation criteria in \citet{hopkins.2013:sf.criteria} and allow star formation to take place only in locally self-gravitating, self-shielding/molecular gas which also exceeds a hydrogen number density threshold $\nc=5\,\cm^{-3}$. Stars form on the local free-fall time when the gas meets these criteria and there is no star formation elsewhere. Once a star forms, it inherits the metallicity of each tracked species from its parent gas particle. Every star particle is treated as a single stellar population with known mass, age, and metallicity, assuming a \citet{kroupa.2002:imf} initial mass function (IMF) from $0.1$--$100\,\Msun$. All the feedback quantities, including ionizing photon budgets, luminosities, supernovae (SNe) rates, mechanical luminosities of stellar winds, etc., are then directly tabulated from the stellar population models in {\sc starburst99} \citep{leitherer.1999:sb99}. We account for several different stellar feedback mechanisms, including (1) local and long-range momentum flux from radiative pressure, (2) energy, momentum, mass and metal injection from SNe and stellar winds, and (3) photo-ionization and photo-electric heating. We follow \citet{wiersma.2009:chemical.enrich} and account for metal production from Type-II SNe, Type-Ia SNe, and stellar winds using the metal yields in \citet{woosley.weaver.1995:sneii.yield}, \citet{iwamoto.1999:snia.yield}, and \citet{izzard.2004:agb.yield}, respectively. The rates of Type-II and Type-Ia SN are separtately computed from {\sc starburst99} and following \citet{mannucci.2006:snia.rates}, respectively. 

We note that the Mg yield from Type II SN in \citet{woosley.weaver.1995:sneii.yield} is $\sim0.4$\,dex lower than typical values in more recent models \citep[e.g.][]{nomoto.2006:sneii.yield}. Therefore, we manually add 0.4\,dex to all Mg abundances in our simulation to compare with observations more accurately. This will have little effect on global galaxy properties, since Mg is not an important coolant (it is simply a ``tracer species''). Also, the total number of Type Ia SNe calculated from \citet{mannucci.2006:snia.rates} is lower than that derived from \citet{maoz.2010:ia.rates} by a factor of a few for a stellar population older than 1\,Gyr; this may lead to predictions of lower Fe, but we cannot simply renormalize the Fe abundances in the simulation. We do not include a sub-resolution metal diffusion model in the simulation; all mixing above the resolution scale is explicitly resolved.

We use the Amiga Halo Finder \citep[{\sc ahf};][]{knollmann.knebe.2009:ahf.code} to identify halos in the simulated box, where we adopt the time-dependent virial overdensity from \citet{bryan.norman.1998:xray.cluster}. In this work, we only study the most massive (hence best-resolved) halo in the zoom-in region, which hosts a disk galaxy of very similar properties to the MW at $z=0$. At each epoch, we define the galactic center at the density peak of most stars and find the stellar half-mass radius as the radius within which the stellar mass equals to a half of the stellar mass within 0.1 virial radius. Then the $Z$-axis is defined to be aligned with the total angular momentum of the gas within 5 stellar half-mass radii. In this paper, we will primarily focus on the stellar component. We do not perform a kinematic decomposition for the stellar content, but take all star particles in the analysis to form an unbiased sample. 

A list of symbols used in this paper and their descriptions are presented in Table \ref{tbl:symbol}. In the rest of this paper, we always mean the $z=0$ age when we quote stellar ages and will predominantly use lookback time ($\tb$) when referring to an epoch in the simulation. In Table \ref{tbl:tb}, we list the conversion between lookback time and redshift at selected epochs for reference.

\begin{figure*}
\centering
\includegraphics[width=\linewidth]{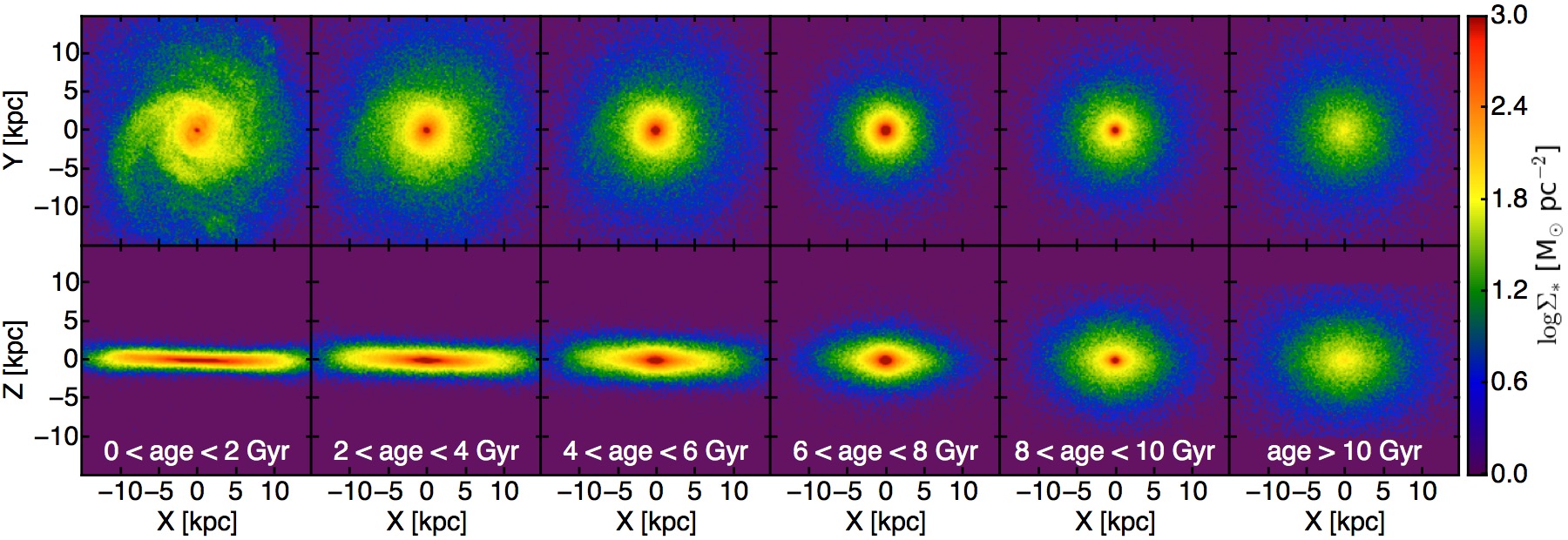} \\
\includegraphics[width=\linewidth]{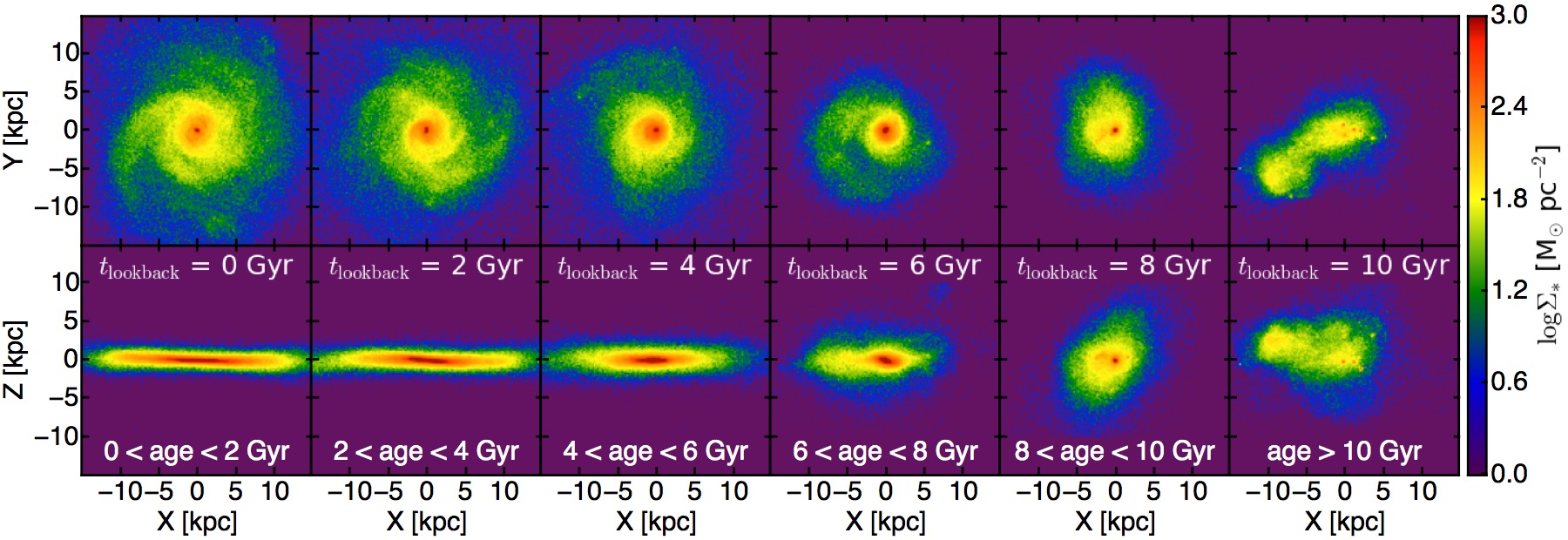} \\
\includegraphics[width=\linewidth]{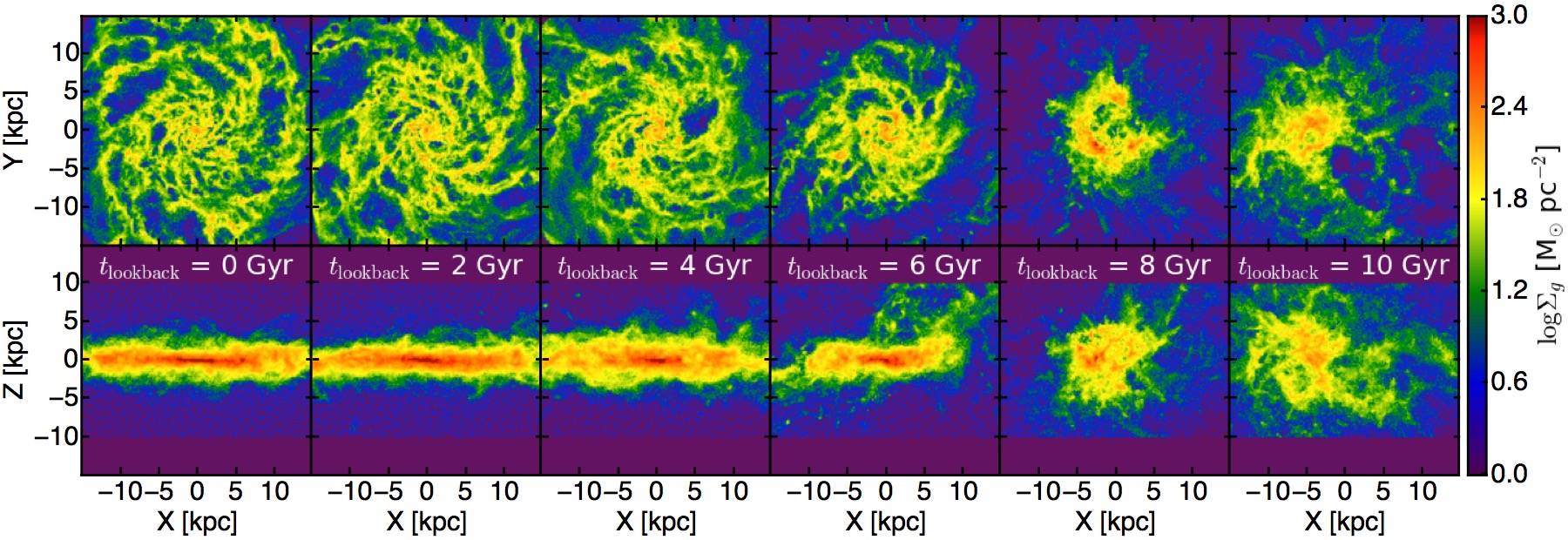} 
\caption{{\em Top:} Morphology of stars in different age intervals at $z=0$. The thickness increases with stellar age, but the scale length first decreases with stellar age in $0<\age<6\,\Gyr$ and then increases in $\age>8\,\Gyr$, leaving stars of $\age\sim6\,\Gyr$ the most radially concentrated (owing to a merger-driven nuclear starburst about this time). {\em Middle:} Morphology of the same stars from each $z=0$ age interval in the top panel, but viewed at the epoch when they just formed (labeled by lookback time) in the galaxy progenitor. Stars younger than 6\,Gyr at $z=0$ were formed in a relatively calm disk. Stars older than 8\,Gyr at $z=0$ were formed in a violent, bursty mode and relax by $z=0$. {\em Bottom:} Morphology of gas, viewed at the same epochs as in the middle panel. At early time, the gas is highly irregular and chaotic. By $\tb\sim6\,\Gyr$ ($z\sim0.7$), the gas eventually formed a disk.}
\label{fig:image}
\end{figure*}

\begin{figure}
\centering
\includegraphics[width=\linewidth]{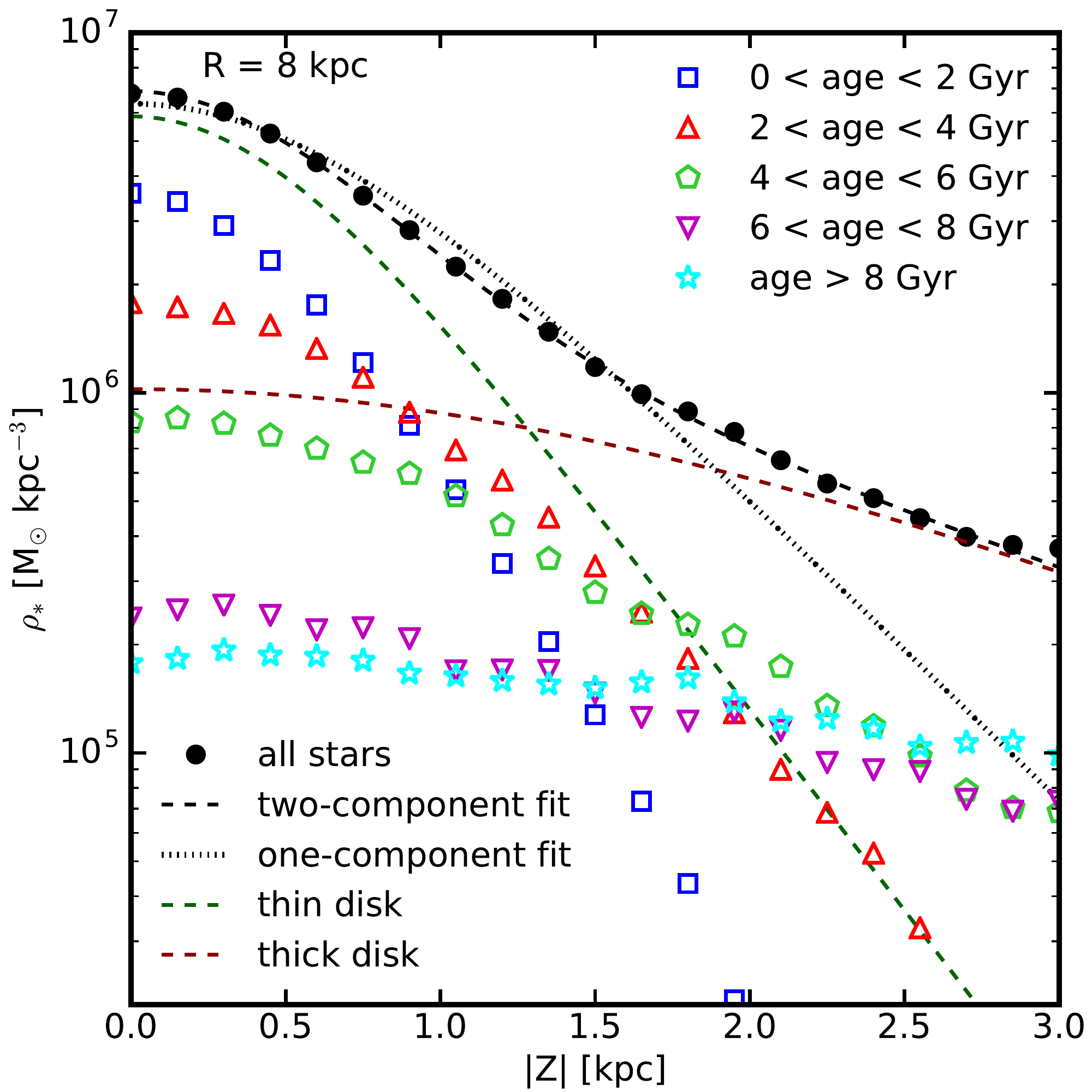}
\caption{Vertical stellar density profile $\rho_{\ast}(\z)$ at $R=8\,\kpc$ (black points) in the simulation at $z=0$. The density profile cannot be well described by a single-component profile (Eqn. \ref{eqn:single}, black dotted line), while a two-component profile provides a good fit (Eqn. \ref{eqn:double}, black dashed line). The dark green and brown dashed lines show the thin- and thick-disk profiles from the fitting. The `thin' and `thick' disks have scale heights of 300\,pc and 1.1\,kpc, respectively, close to the MW scale heights around the solar neighborhood \citep[300\,pc and 1450\,pc, e.g.][]{gilmore.reid.1983:thick.disk}. The `thick disk' contributes 36\% of the total stellar mass around $R=8\,\kpc$. Note that the stellar densities, thin- and thick-disk scale heights, and the mass fraction of the thick disk in our simulation are in good agreement with observations and other simulations in the literature. The open symbols show the density of stars in different age intervals. Stars younger than 4\,Gyr contribute more than 90\% of the mass in the `thin disk', while the `thick disk' is made almost entirely of stars older than 4\,Gyr.}
\label{fig:profile}
\end{figure}

\section{Results}
\label{sec:results}
\subsection{General Picture}
At high redshifts, the galaxy accretes gas rapidly and undergones multiple mergers, producing violent, bursty star formation, until a final minor merger finished at $z\sim0.7$ (corresponding to a look-back time of $\tb\sim6\,\Gyr$). Since then, a calm, stable gas disk was formed and maintained, with stars forming in the disk at a nearly constant rate ($\sim7\,\Msun\,\yr^{-1}$, integrated across the entire disk) regulated by stellar feedback.

The top panel in Fig. \ref{fig:image} illustrates the stellar morphologies at $z=0$ for stars in the galaxy in six different $z=0$ age intervals. The top and bottom panels show the stellar surface density viewed face-on and edge-on, respectively. The thickness increases with stellar age, from a thin disk-like structure to more  spheroidal morphology, broadly consistent with the MW \citep{bovy.2012:mono.structure} and other simulations \citep[e.g.][]{brook.2012:thin.thick.disk.halo,bird.2013:disk.galaxy.assembly,stinson.2013:magicc.mw.disk,minchev.2013:chemodynamical,martig.2014:disk.map.structure}. On the other hand, the radial morphology first shrinks with increasing age (`inside-out' growth), but then becomes less concentrated for ages greater than 8\,Gyr, leaving intermediate-age stars ($\age\sim6\,\Gyr$) the most radially concentrated. This is in contrast with the results in \citet{bovy.2012:mono.structure} and other simulations where the scale length decreases monotonically with stellar age (oldest stars have the smallest scale lengths). This directly owes to a minor merger in the simulation around lookback time $\tb\sim6\,\Gyr$ ($z\sim0.7$), which drove a concentrated nuclear starburst. Afterwards, the disk formed inside out.

The middle panel in Fig. \ref{fig:image} shows the stellar morphologies for the same stars shown in the top panel (divided into the same $z=0$ age intervals), but viewed at the epoch when they just formed (labeled by look-back time). In other words, we trace the galaxy back to these epochs, and show the young stars in the main progenitor galaxy at that time. Stars older than 8\,Gyr were born to be a chaotic, non-disk-like structure. For illustrative purposes, we also show gas morphologies at the same epochs in the bottom panel in Fig. \ref{fig:image}. During the early stage of galaxy assembly when the stellar mass was sufficiently low, this galaxy experienced bursty, chaotic star formation \citep{sparre.2017:fire.sf.burst}. Starbursts drive bursts of gas outflows with high efficiency \citep{muratov.2015:fire.mass.loading}, and the bursty outflows in turn modify the potential and cause radial migration of stars, resulting in radial expansion and quasi-spherical morphology for stars older than 8\,Gyr \citep{elbadry.2016:fire.migration}. A gas disk is formed by $\tb\sim6$\,Gyr ($z\sim0.7$). Below $\tb\lesssim6$\,Gyr, star formation takes place in a relatively calm mode, with stars forming in a relatively stable disk at a rate self-regulated by feedback, and there are no longer large scale outflows \citep{muratov.2015:fire.mass.loading,daa.2016:fire.baryon.cycle}. \citet{hayward.hopkins.2017:wind.driving} proposed an analytic model and argued that such bursty-to-calm transition is expected in massive galaxies at late times, due to the change of ISM structure at low gas fractions.

We estimate the fraction of stars that comes from mergers or tidally disrupted satellites, i.e. stars formed outside the main progenitor galaxy, using the particle tracking technique developed by and presented in \citet{daa.2016:fire.baryon.cycle}. We find that only $\lesssim10\%$ of the stellar mass in the $z=0$ galaxy was formed {\em ex situ} and this contribution is only significant far above the galactic plane ($\z\gtrsim5\,\kpc$). For example, during the last minor merger at $\tb\sim6\,\Gyr$ ($z\sim0.7$), the passing-by satellite has been tidally disrupted and its stars are re-distributed in the diffuse halo. Within the galactic disk, which we select to be in galactocentric radius $R=4$--14\,kpc and $\z<3\,\kpc$ (to exclude bulge and halo stars), stars that were formed {\em ex situ} contribute no more than a few percent of the stellar mass, so we will ignore them in the analysis below.

\begin{figure}
\centering
\includegraphics[width=\linewidth]{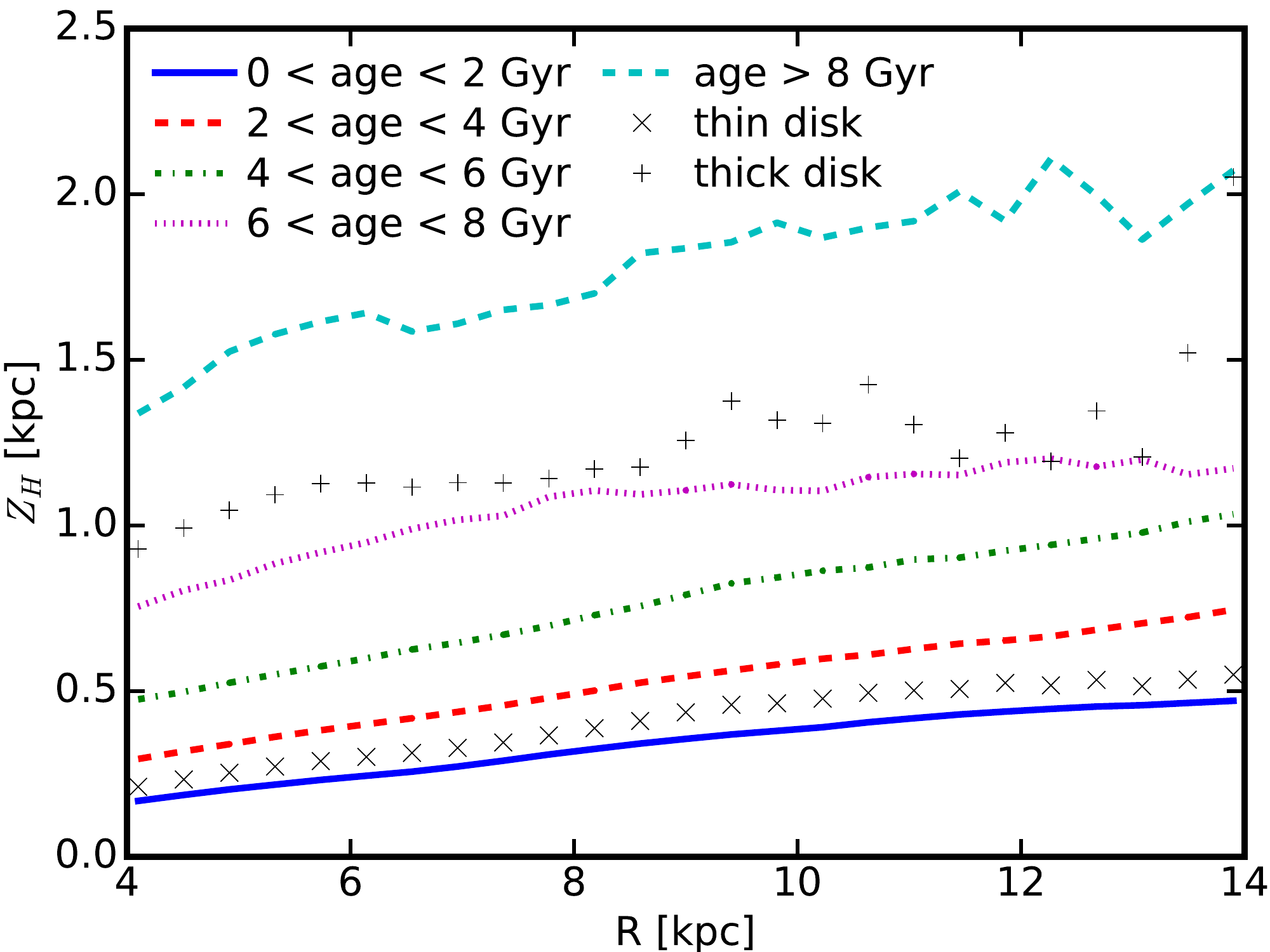}
\caption{Scale heights of stars in five age intervals, as a function of galactocentric radius in the disk. Each population can be well-described by a single-component profile from Equation (\ref{eqn:single}). At fixed radius, $\zh$ increases with stellar age. For a given population, $\zh$ increases with $R$. The cross and plus symbols show the scale heights of the thin- and thick-disk component, respectively. At all radii, the `thin disk' has a scale height close to that of stars younger than 4\,Gyr.}
\label{fig:ZHZ0}
\end{figure}

\subsection{Disk Structure}
\label{sec:disk}
One common argument for the presence of a thick disk in the MW and nearby disk galaxies is that the stellar density profile along the $\z$-direction cannot be well-described by a single-component profile
\be
\label{eqn:single}
\rho_{\ast}(\z) = \rho_{\ast}(0) \, \sech^2 \left( \frac{\z}{2\zh} \right),
\ee
where $\zh$ is the scale height, but requires a second component 
\be
\label{eqn:double}
\rho_{\ast}(\z)  =  \rho_{\ast,1}(0) \, \sech^2 \left( \frac{\z}{2Z_{H,1}} \right)
	 + \rho_{\ast,2}(0) \, \sech^2 \left( \frac{\z}{2Z_{H,2}} \right)
\ee
\citep[e.g.][]{gilmore.reid.1983:thick.disk,yoachim.dalcan.2006:thick.disk,comeron.2011:thick.disk.ngc.4244,comeron.2012:thick.disk.s4g}. In Fig. \ref{fig:profile}, we show the vertical stellar density profile $\rho_{\ast}(\z)$ at $R\sim8\,\kpc$ in our simulation (black dots) at $z=0$ and fit it with a single-component profile (grey dotted line) and a two-component profile (black dashed line), respectively. The dark green and brown lines show the thin- and thick-components, respectively.

We find that a two-component profile provides a much better fit than a single-component profile. The two components, which we refer as the `thin' and `thick' disks, have scale heights of $Z_{H,1}\sim300$\,pc and $Z_{H,2}\sim1.1\,\kpc$, respectively, close to the observed MW-disk scale heights around the solar neighborhood \citep[300\,pc and 1450\,pc, e.g.][]{gilmore.reid.1983:thick.disk}. Derived from the profile fitting, the thick disk component contributes 36\% of the stellar mass at $R=8\,\kpc$, broadly in agreement with measurements of nearby edge-on disk galaxies \citep[e.g.][]{yoachim.dalcan.2006:thick.disk} as well as other simulations \citep[e.g.][]{brook.2012:thin.thick.disk.halo,bird.2013:disk.galaxy.assembly,minchev.2013:chemodynamical}. Note that this is far greater than the fraction of stars that were formed {\em ex situ}, which is $\lesssim5\%$ at $R=8\,\kpc$, so the `thick disk' in our simulation does not originate from accreted satellite galaxies.

In Fig. \ref{fig:profile}, we further decompose the density profile into five bins according to stellar age (open symbols). Qualitatively, the thickness of stars increases with age, with youngest stars being most concentrated to the mid-plane and the oldest stars being most vertically extended. This is consistent with the visualization shown in Fig. \ref{fig:image}. We find that over 90\% of the mass in the `thin disk' is contributed by stars younger than 4\,Gyr, while the `thick disk' is made of stars older than 4\,Gyr. Note that our thin-to-thick disk decomposition is purely based on the mass density at this point. In Section \ref{sec:thick}, we will further discuss the formation mechanisms of both components. 

Stars in each age interval in Fig. \ref{fig:profile} can be well-described by a single-component profile from Equation \ref{eqn:single} \citep[see also, e.g.][]{bird.2013:disk.galaxy.assembly,martig.2014:disk.map.structure,minchev.2013:chemodynamical,minchev.2015:thick.disk.formation}. In Fig. \ref{fig:ZHZ0}, we further show $\zh$ as a function of $R$ for stars in all five age intervals. Only $R=4$--14\,kpc is shown to minimize contamination from the bulge component, which is important within $R<4\,\kpc$. Stars older than 8\,Gyr have very large scale heights ($\zh>1.5$\,kpc), since they were formed during the chaotic phase and have relaxed to be quasi-spherical by $z=0$. Stars younger than 6\,Gyr have considerably smaller scale heights, since they were formed in a disk. Even for these stars, the scale heights increase with stellar age at any radius. For example, the scale heights of stars with age 4--6\,Gyr are larger than those of stars with $\age<2$\,Gyr by a factor of 2. Note that this is equivalent to the observed age--velocity dispersion relation \citep[e.g.][]{nordstrom.2004:age.metal.kin}, since the vertical velocity dispersion is proportional to disk thickness as expected from disk equilibrium. For comparison, we also show the scale heights of the `thin' and `thick' disks as a function of galactocentric radius (black cross and plus symbols). At all radii, the `thin disk' scale heights are comparable to those of stars younger than 4\,Gyr ($Z_{H,1}=200$--500\,pc), while the `thick disk' represents a median stellar age of 8\,Gyr ($Z_{H,2}=1$--1.5\,kpc). Moreover, the disk is flaring for stars younger than 6\,Gyr -- the scale height increases with $R$, with $\zh$ being a factor of 2 larger at $R=14$\,kpc than that at $R=4$\,kpc. The disk flaring broadly agrees with observations in the MW stellar disk \citep[e.g.][]{momany.2006:mw.warp.flare,kalberla.2014:mw.disk.flare,lopez.2014:mw.disk.flare,bovy.2016:mw.star.structure}.

\begin{figure}
\centering
\includegraphics[width=\linewidth]{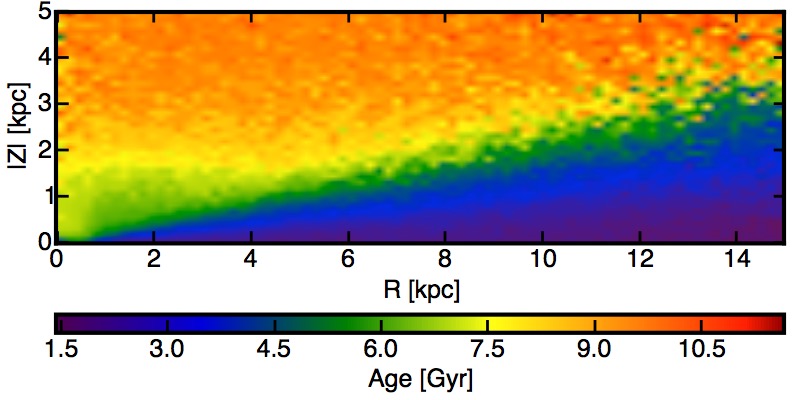} \\
\includegraphics[width=\linewidth]{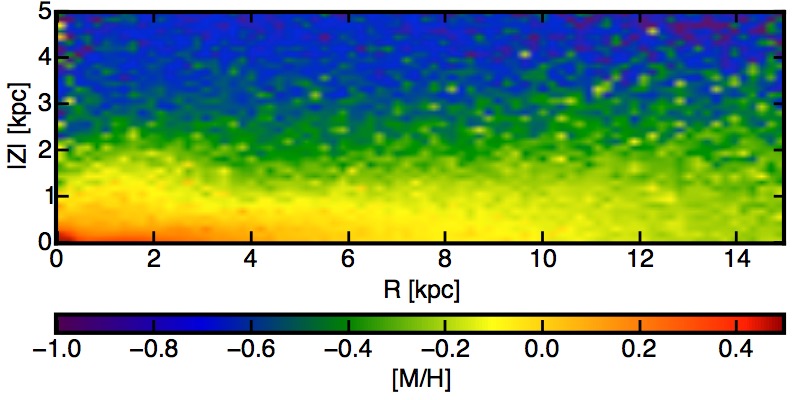} 
\caption{{\em Top:} Median stellar age as a function of $R$ and $\z$. The median stellar age naturally follows the disk structure. At fixed radius, stellar age increases with $\z$. Above the mid-plane, the stellar age decreases with $R$ due to disk flaring. {\em Bottom:} Stellar metallicity in the disk. $\MH$ is higher at the inner disk and near the mid-plane than at the outer disk and large heights.}
\label{fig:age}
\end{figure}

\begin{figure*}
\centering
\includegraphics[width=\linewidth]{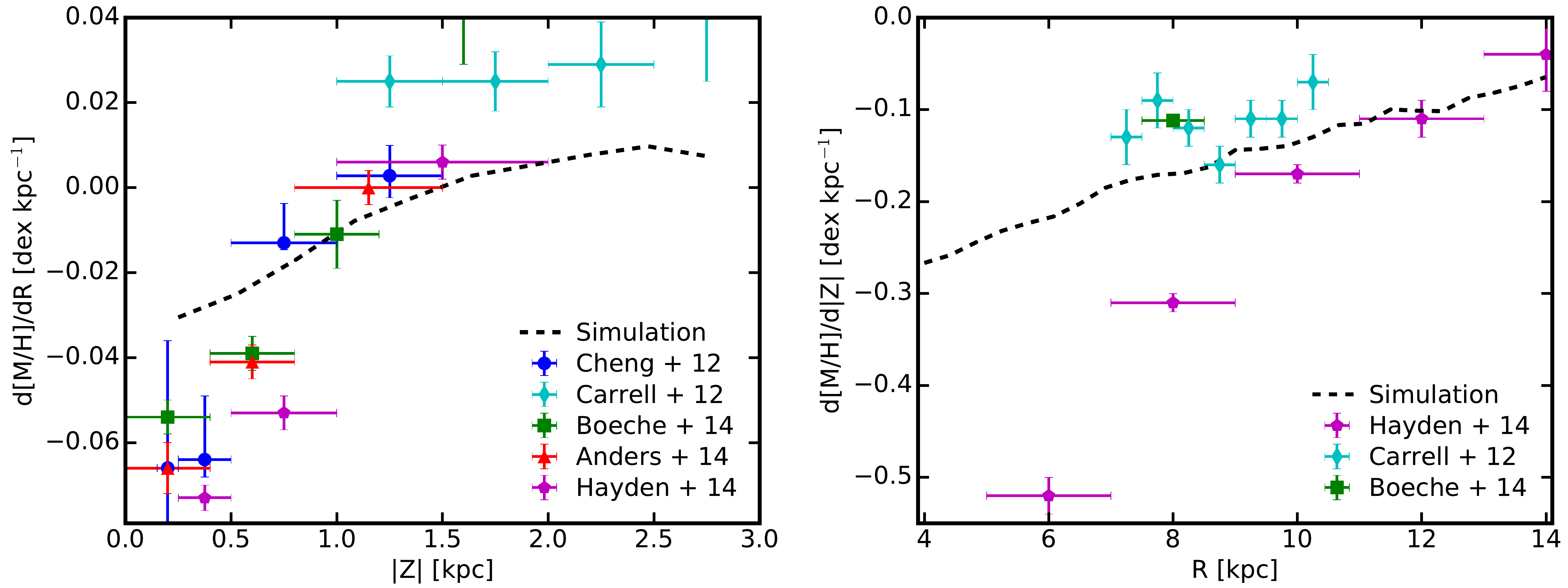}
\caption{{\em Left}: Radial metallicity gradient $\rgrad$ as a function of height $\z$. {\em Right}: Vertical metallicity gradient $\vgrad$ as a function of galactocentric radius $R$. $\rgrad$ is negative at the mid-plane, gradually increases with $\z$, and eventually becomes positive above $\z\sim1.5\,\kpc$. $\vgrad$ is negative at all radii, but is stronger at smaller radii. These trends are qualitatively consistent with observations in the MW disk. A number of observations are shown, including \citet{cheng.2012:segue.stellar.grad}, \citet{carrell.2012:metal.grad.disk.dwarf}, \citet{boeche.2014:rave.giant.grad}, \citet{anders.2014:chemo.apogee.dr1}, and \citet{hayden.2014:chemo.apogee}.} 
\label{fig:ZGrad}
\end{figure*}

\begin{figure*}
\centering
\includegraphics[width=\linewidth]{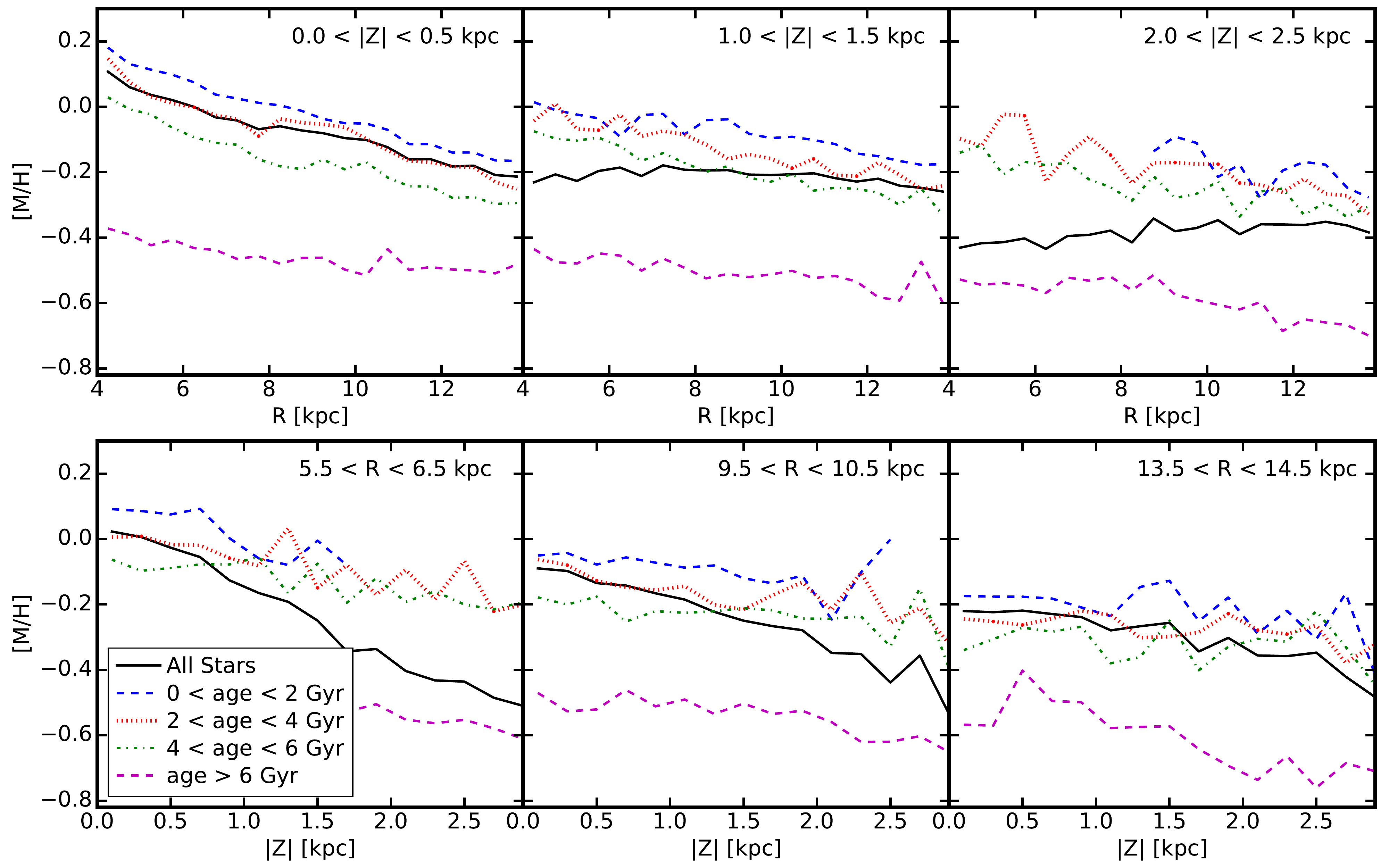}
\caption{{\em Top}: Radial metallicity profile in layers with $\z=0$--0.5, 1.0--1.5, and 2.0--2.5\,kpc. {\em Bottom}: Vertical metallicity profile at radii $R=6,\,10,\,14\,\kpc$. We show the metallicity profiles for all stars (black solid lines) as well as in bins of different stellar ages. The flattening and inversion of the radial metallicity gradient at high $\z$ follows the negative age gradient at these heights. The steepening of the vertical metallicity gradient at smaller radii results from a stronger age gradient. The stellar age gradient is a natural consequence of disk structure. These results are in line with the predictions in \citet[][fig. 10]{minchev.2014:chemodynamical}.}
\label{fig:ZGradAge}
\end{figure*}

\begin{figure*}
\centering
\includegraphics[width=\linewidth]{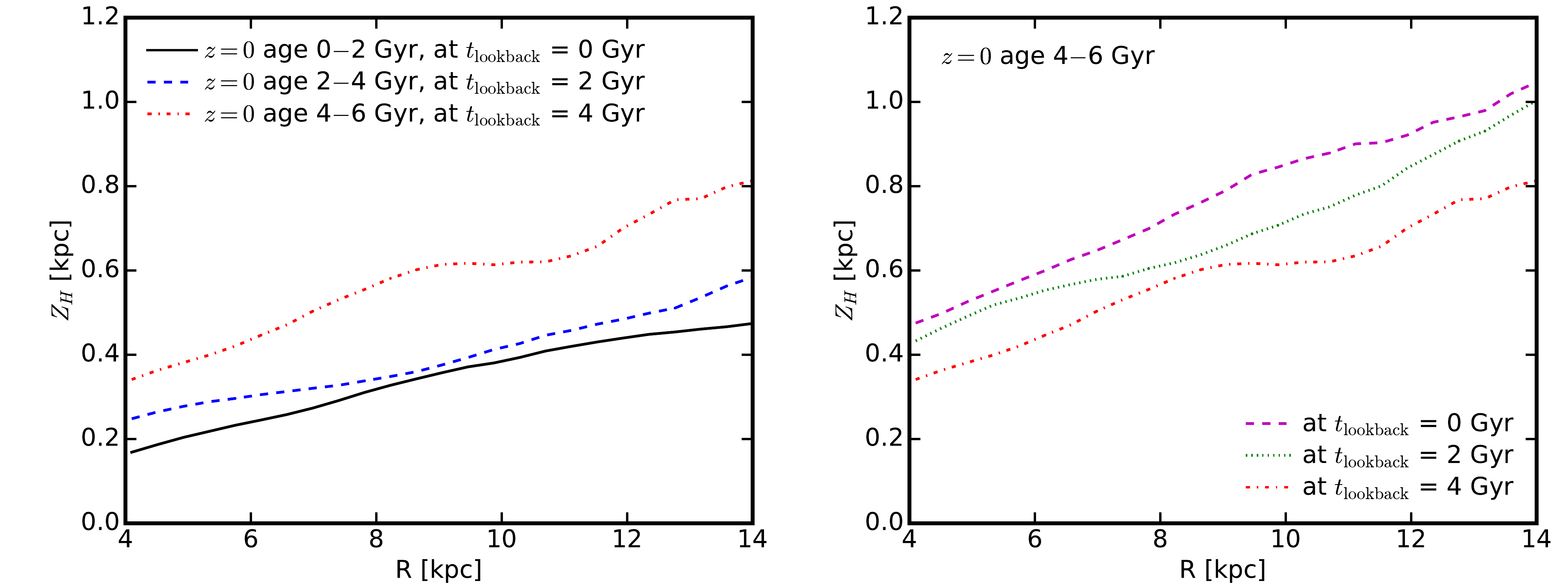}
\caption{{\em Left}: Scale heights of stars in three $z=0$ age intervals (0--2, 2--4, 4--6\,Gyr), measured at the epoch when the stars formed (labeled by lookback time). {\em Right}: The evolution of scale heights for stars of age 4--6\,Gyr at $z=0$, as they would be observed at three different epochs (labeled by lookback time) after their formation. Stars formed at an early epoch were born in a thicker disk than those formed at later times, because the star-forming disk was more gas-rich and therefore turbulent. The scale heights of those populations then increase with time due to kinematic heating.}
\label{fig:ZHEvolve}
\end{figure*}

\subsection{Age and Metallicity Gradients}
\label{sec:zgrad}
In Fig. \ref{fig:age}, we show the median stellar age as a function of galactocentric radius $R$ and height $\z$ (upper panel) for $R=0$--15\,kpc and $\z=0$--5\,kpc in our simulation at $z=0$. At each radius, the median stellar age increases with $\z$, resulting in a significant positive vertical age gradient in the disk. Moreover, there is also a moderate negative radial age gradient above the mid-plane, as the median stellar age decreases with $R$ at fixed $\z$. These features naturally follow the disk structure presented above: (1) differential scale heights of stars in different age intervals and (2) the disk flaring for any single-age stellar population. These results are in line with predictions from \citet{minchev.2015:thick.disk.formation} and observations from \citet{martig.2016:age.gradient.mw.thick}. In Fig. \ref{fig:age}, we also show the mass-weighted mean stellar metallicity as a function of $R$ and $\z$ (bottom panel). Qualitatively, the stellar metallicity is higher at the inner disk and near the mid-plane than at the outer disk and large heights.

In Fig. \ref{fig:ZGrad}, we further show the radial metallicity gradient $\rgrad$ as a function of height $\z$ (left panel) and vertical metallicity gradient $\vgrad$ as a function of galactocentric radius $R$ (right panel). The black dashed lines show the values measured in the $z=0$ snapshot of our simulation. The radial metallicity gradient at a certain $\z$ is measured using stars in a layer of thickness $\Delta\z=0.5\,\kpc$ and by fitting the radial metallicity profile from $R=4$--14\,kpc with a simple linear function. For vertical metallicity gradients, we use stars in annuli of $\Delta R=1\,\kpc$ and fit the vertical metallicity profile from $\z=0$--2.5\,kpc with a linear function. We take all star particles into account. 

In Fig. \ref{fig:ZGrad}, we also compare our results with published radial and vertical metallicity gradients measured from different samples of MW stars in the literature \citep{cheng.2012:segue.stellar.grad,carrell.2012:metal.grad.disk.dwarf,boeche.2014:rave.giant.grad,anders.2014:chemo.apogee.dr1,hayden.2014:chemo.apogee}. The horizontal error bars show the $R$ or $\z$ interval where the metallicity gradient is measured. Our simulation is qualitatively consistent with observations, despite the fact that the slopes are not identical -- $\rgrad=-0.03\,\dkpc$ at $\z<0.5\,\kpc$ is shallower than the canonical MW value of $-0.06\,\dkpc$, but is close or slightly steeper than that of $-0.02\,\dkpc$ in M31 \citep[e.g.][]{gregersen.2015:m31.metal.grad}. It is difficult to know whether this discrepancy is meaningful, without a large, statistically significant sample (both simulated and observed). We find that the radial gradient is negative and steepest near the mid-plane, gradually flattens, and finally becomes positive above $\z\gtrsim1.5\,\kpc$, as observed in the MW \citep[e.g.][]{cheng.2012:segue.stellar.grad,boeche.2014:rave.giant.grad,anders.2014:chemo.apogee.dr1} and predicted in other simulations \citep[e.g.][]{minchev.2014:chemodynamical,minchev.2015:thick.disk.formation}. The vertical gradient is negative at any radius between $R=4$--14\,kpc, but is steeper at inner radii. This trend is qualitatively consistent with observations in \citet{hayden.2014:chemo.apogee}.

To understand why the metallicity gradients have such behavior, in Fig. \ref{fig:ZGradAge}, we break down the radial (top panels) and vertical (bottom panels) metallicity profiles into four age intervals: $\age<2\,\Gyr$ (blue dashed lines), $2<\age<4\,\Gyr$ (red dotted lines), $4<\age<6\,\Gyr$ (green dash-dotted lines), and $\age>6\,\Gyr$ (magenta dashed lines). The black lines show the metallicity profile of all stars. To leading order, metallicity is a proxy of stellar age, with young stars being more metal-enriched than old stars (also see Section \ref{sec:mono}). The top panels in Fig. \ref{fig:ZGradAge} show the radial metallicity profiles from $R=4$--14\,kpc in three layers: $0<\z<0.5\,\kpc$, $1.0<\z<1.5\,\kpc$, and $2.0<\z<2.5\,\kpc$. The flattening and inversion of the radial metallicity gradient can be naturally understood from the negative age gradient at large heights. For example, in the $2.0<\z<2.5\,\kpc$ layer, old, metal-poor stars dominate at $R=4$ (where the disk scale heights of young stars are small, i.e. $\zh\sim0.3$\,kpc), while younger, more metal-enriched stars take over at much larger radius (where the young stellar disk is thicker in absolute units, e.g. $\zh\sim1$\,kpc at $R=14\,\kpc$). This leads to an overall positive radial metallicity gradient at this fixed height. Note that our interpretations here agree well with the chemo-dynamical model in \citet[][fig. 10]{minchev.2014:chemodynamical}. The bottom panels in Fig. \ref{fig:ZGradAge} show the vertical metallicity profile at $R=6$, 10, and 14\,kpc, respectively. The reasons why the vertical metallicity gradient flattens at large radius are twofold: (1) the metallicity of young stars is lower at larger radius than at small radius and (2) the age gradient is much weaker at larger radius.

The negative radial metallicity gradient for stars younger than 6\,Gyr (those formed in a disk) near the mid-plane is inherited from the parent star-forming gas disk \citep{ma.2017:fire.metallicity.gradient}. A negative radial metallicity gradient is expected from the coevolution between the gas disk and stellar disk \citep[e.g.][]{ho.2015:metal.grad.coevolve}. In short, suppose a pure gas disk has formed with an exponential surface density profile and begun to form stars. The star formation efficiency is higher at the inner disk than at the outer disk according to the observed Kennicutt--Schmidt law \citep{kennicutt.1998:sf.law.review}. Under the local closed-box assumption, the inner disk would be enriched more rapidly than the outer disk, leading to a negative radial metallicity gradient in the disk. Nonetheless, the slope of such a gradient can be affected by disk scale length, radial inflow and mixing, disk pre-enrichment, etc. A comprehensive analysis of the radial metallicity gradient and its dependence on galaxy properties with a larger sample of simulations is presented in a companion study \citep{ma.2017:fire.metallicity.gradient}.

\subsection{Dynamical Evolution of the Stellar Disk}
\label{sec:evolve}
In Section \ref{sec:disk}, we show that even for stars that are initially formed in a disk (i.e. stars younger than $\age\sim6\,\Gyr$ by $z=0$), by $z=0$, their scale height increases with stellar age at all radii (Fig. \ref{fig:ZHZ0}). To explain this, we first explore the scale heights of stars at the time when they just formed. In the left panel in Fig. \ref{fig:ZHEvolve}, we show the scale heights for stars in three $z=0$ age intervals: 0--2\,Gyr, 2--4\,Gyr, and 4--6\,Gyr, but measured just after their formation time (labeled by lookback time). In other words, these stars are younger than 2\,Gyr at the time we measure their scale heights. The newly formed stars inherit the scale heights and velocity dispersion from the cold star-forming gas in the disk where they were born. In general, stars formed earlier (which are older today) were born with larger scale heights than stars formed at late times. For example, stars with $z=0$ age 4--6\,Gyr (formation time at $\tb=4\,\Gyr$) were born with a scale height of $\zh\sim0.4$ (0.8)\,kpc at $R=4$ (14)\,kpc at this epoch, larger by a factor of $\sim2$ compared to the `at birth' scale heights of stars formed at $z=0$. This naturally follows the evolution of the gas disk, because the thickness of a star-forming disk, where self-regulation by feedback yields a Toomre parameter $Q\sim1$, is proportional to its gas fraction \citep[e.g.][]{thompson.2005:starburst.disk,faucher.2013:self.regulation}, which is higher at early times.

We now examine how the scale height evolves over time. In the right panel in Fig. \ref{fig:ZHEvolve}, we show the scale heights at three post-formation epochs (labeled by lookback time), for stars in the $z=0$ age interval 4--6\,Gyr. At all radii, the scale height increases by $\sim30\%$ (or $\sim0.2\,\kpc$ in absolute units) over the 4\,Gyr from their formation time to $z=0$. During the same period, the vertical velocity dispersion has also increased consistently. Our simulation shows that kinematic heating plays a non-negligible role in the formation of the thick disk, in line with the predictions in \citet{minchev.2013:chemodynamical} and \citet{martig.2014:disk.age.velocity}, but in contrast with the argument in \citet{bird.2013:disk.galaxy.assembly}.

Several mechanisms have been proposed to cause such kinematic heating, including: (1) bars and spiral arms \citep[e.g.][]{sellwood.carlberg.1984:spiral,minchev.quillen.2006:spiral,saha.2010:bar.heating,faure.2014:vertical.heating.spiral,yurin.springel.2015:stellar.disk,grand.2016:bar.heating}, (2) radial migration (e.g. \citealt{schonrich.binney.2009:structure,loebman.2011:thick.migration}; however, see e.g. \citealt{minchev.2012:migration.no.heat,vera-ciro.2014:radial.migration,grand.2016:bar.heating}), (3) perturbation of satellites and sub-halos \citep[e.g.][]{quinn.1993:merger.heating,kazantzidis.2008:thick.disk.merger,purcell.2009:disk.heating.merger,gomez.2013:satellite.heating}, and (4) scattering by giant molecular clouds (GMCs) or star clusters \citep[e.g.][]{spitzer.schwarz.1951:cloud.scatter,spitzer.schwarz.1953:cloud.scatter,aumer.2016:disk.heat.gmc.spiral}. In a cosmological context, these mechanisms are usually combined and difficult to isolate in practice. For example, gravitational perturbation of satellites can induce bars and spiral arms \citep[e.g.][]{purcell.2011:nature.dwarf.infall}, which further result in kinematic heating and radial migration \citep{lynden-bell.kalnajs.1972:spiral}. Scattering by massive GMCs is also needed to redistribute the energy between planar and vertical motions \citep{carlberg.1987:gmc.scatter}. In our simulation, the increase of disk thickness and velocity dispersion is roughly a linear function of time, indicating that spiral arms may be the dominant heating mechanism, as suggested by an analysis of a large sample of disk galaxy simulations \citep{grand.2016:bar.heating}.

The flaring of the stellar disk is present `at birth' and preserved during kinematic heating. At their formation time, stars inherited the flaring of their parent gas disk, which is likely to be a natural consequence of hydrostatic equilibrium in a galactic potential \citep[e.g.][]{olling.1995:disk.flare.potential,obrien.2010:halo.profile.disk.flare,allaert.2015:gas.disk.structure}, although disk flaring may also be induced and enhanced by mergers \citep[e.g.][]{bournaud.2009:thick.disk.form,purcell.2011:nature.dwarf.infall} and radial migration \citep[e.g.][]{minchev.2012:migration.no.heat}.

\begin{figure}
\centering
\includegraphics[width=\linewidth]{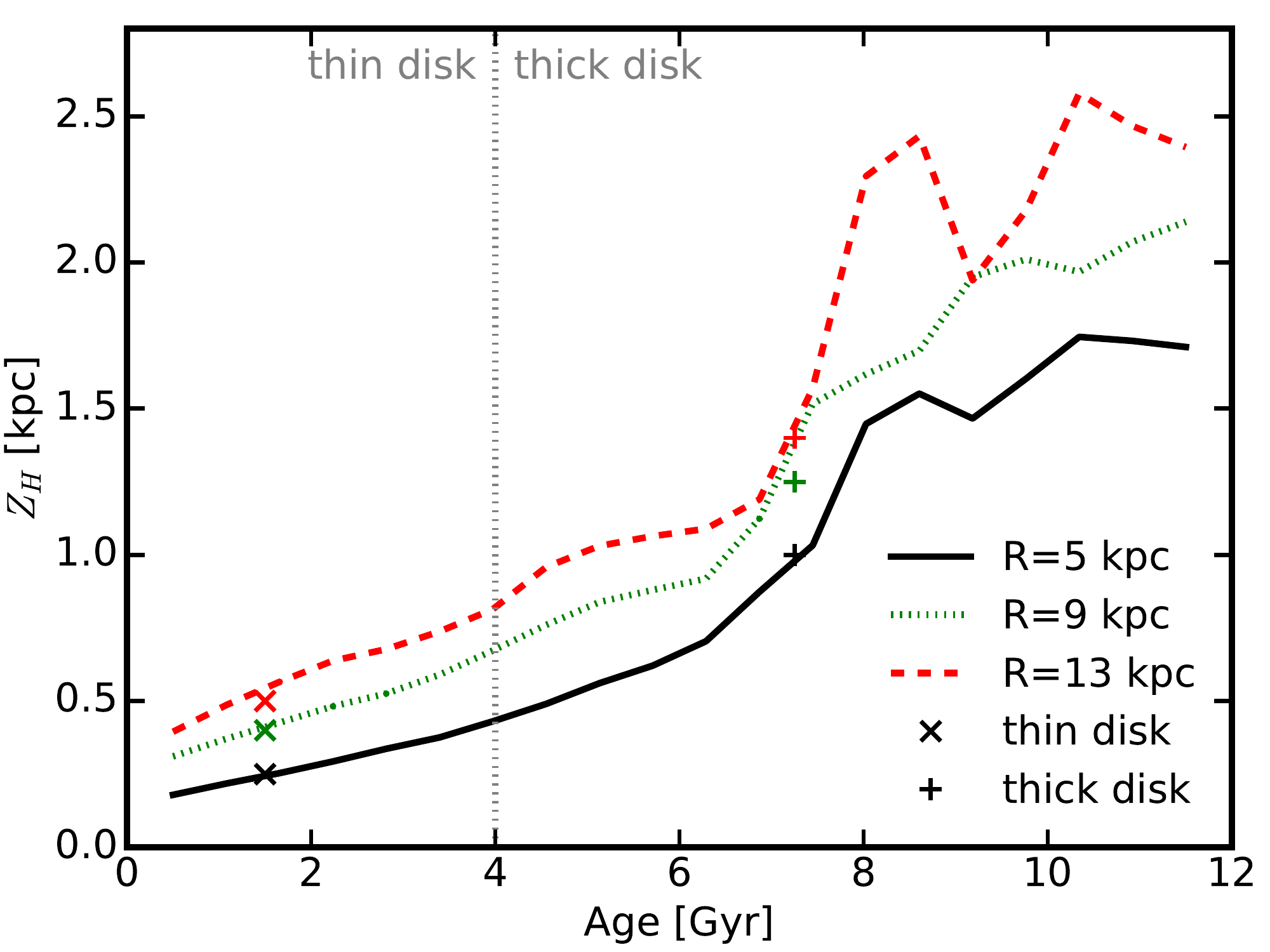}
\caption{Scale heights as a function of stellar age at $R=5$, 9, and 13\,kpc. The cross and plus symbols show the scale heights of the `thin' and `thick' components obtained from the profile fitting (e.g. see Fig. \ref{fig:profile}) at three radii, but we intentionally select their $x$-coordinates to match the curve. At all radii, the scale heights increase dramatically above stellar age 6\,Gyr, since these stars were formed in a chaotic, bursty mode. In terms of mass density, the thin and thick disks can be separated by stars younger and older than 4\,Gyr, as illustrated by the vertical dotted line. The thick disk defined in this way contains two distinct populations of stars: (1) stars older than 6\,Gyr which were formed in the chaotic, bursty mode and (2) stars in 4--6\,Gyr age interval that were formed in a gas-rich disk and kinematically heated after their formation. Note that the gas disk evolved smoothly during the past 6\,Gyr, so there is no sharp transition at 4\,Gyr ago when the thin disk at $z=0$ started to form.}
\label{fig:ZHAgeRad}
\end{figure}

\begin{figure*}
\centering
\begin{center}
\begin{tabular}{cc}
\includegraphics[width=0.49\linewidth]{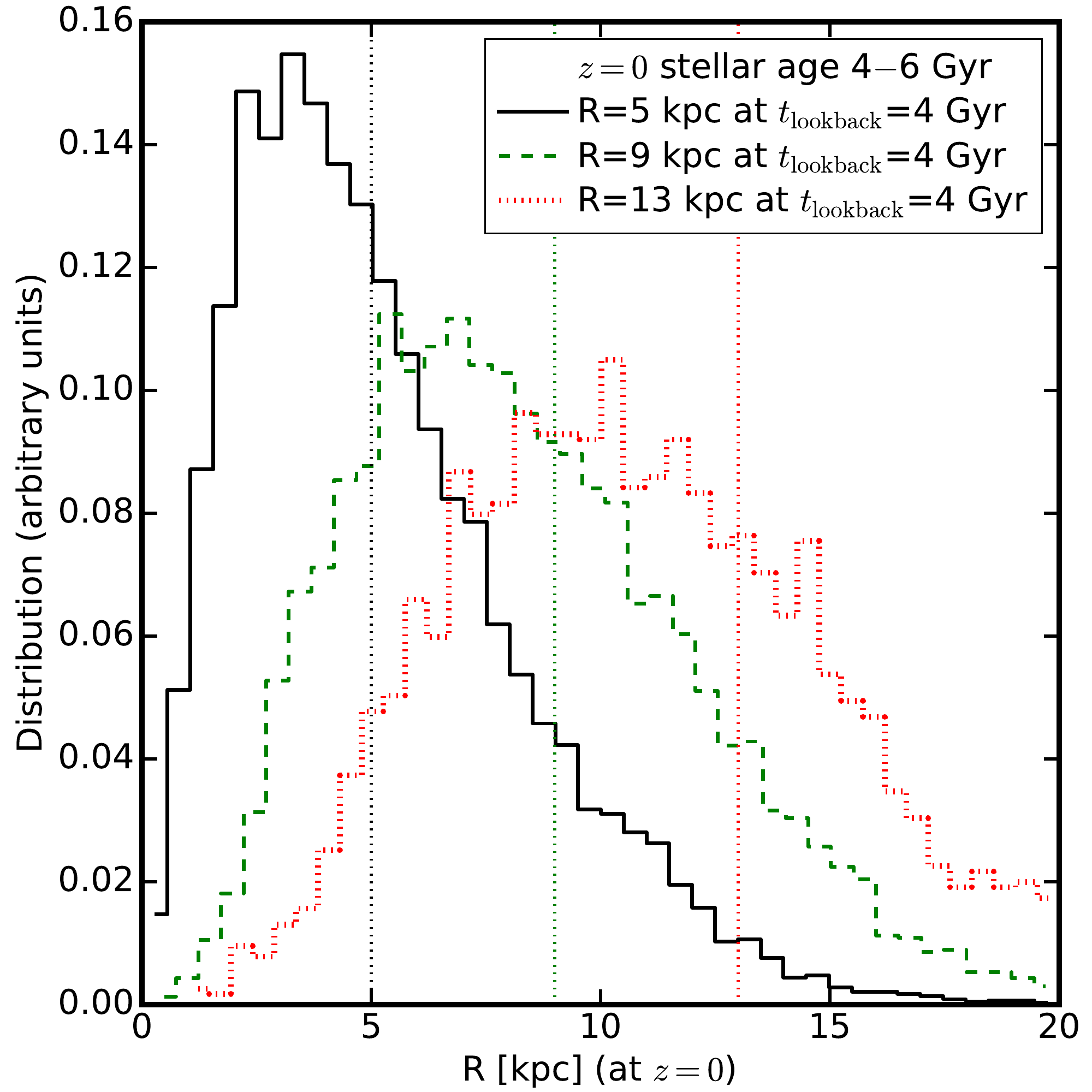} &
\includegraphics[width=0.49\linewidth]{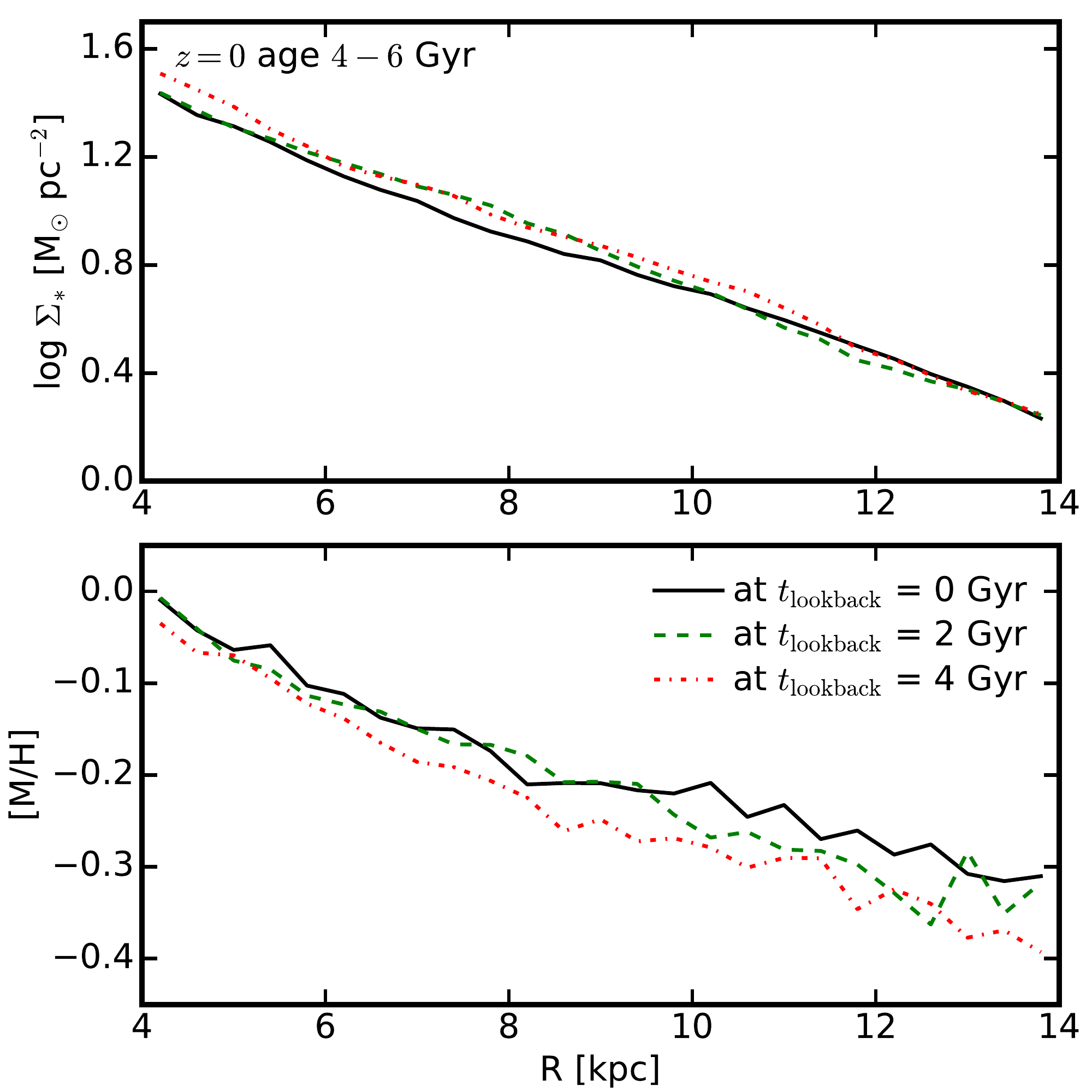}
\end{tabular}
\end{center}
\caption{{\em Left:} Distribution of galactocentric radius at $z=0$ for stars in the age interval 4--6\,Gyr that were located in three annuli centered at $R=5$, 9, and 13\,kpc (of 1\,kpc width) near the disk mid-plane ($\z<0.5\,\kpc$) at $\tb=4\,\Gyr$. Stars in a given annulus 4\,Gyr ago have a wide distribution in $R$ by $z=0$. A small fraction (less than 10\%) of stars have migrated to very large radii ($\Delta R>5\,\kpc$), while more than half of the stars have moved inward ($\Delta R<0$), as expected from exchange of angular momentum. {\rm Right:} Stellar surface density (top) and metallicity (bottom) profiles from $R=4$--14\,kpc for all stars with $z=0$ age 4--6\,Gyr, but measured at three epochs ($\tb=0$, 2, and 4\,kpc). The surface density does not change more than 0.05\,dex, while the average stellar metallicity slightly increased at large radii, but the net effect of stellar migration on the properties of a galactic disk is weak at late times, in line with the results from other simulations \citep[e.g.][]{minchev.2013:chemodynamical,grand.kawata.2016:radial.migration}.}
\label{fig:migration}
\end{figure*}

\section{Discussion}
\label{sec:discussion}
In this section, we discuss some observational and theoretical implications of our simulation. Although our analysis in Section \ref{sec:results} is based on a single simulation, preliminary analysis of several other MW-mass disk galaxy simulations indicates that the disk structure and dynamic evolution are similar in other systems, including one using 8 times higher mass resolution from \citet[][see Appendix \ref{apd:res} for more details]{wetzel.2016:high.res.mw.letter}, despite the fact that these simulations are run with a different hydrodynamic method and slightly modified numerical implementations of the feedback model (the FIRE-2 code, see Hopkins et al., in preparation). This suggests that the results presented in Section \ref{sec:results} are typical in similar systems and insensitive to resolution and numerical method, as implied also by the good agreement between our results and many other simulations \citep[e.g.][]{brook.2012:thin.thick.disk.halo,minchev.2013:chemodynamical,minchev.2014:chemodynamical,martig.2014:disk.map.structure,martig.2014:disk.age.velocity}. This is expected since our key results are derived from global processes and can be understood with simple analytic models, including (1) star formation is bursty at high redshift and becomes relatively stable at late times, (2) the thickness of the star-forming gas disk decreases at low gas fraction, and (3) the kinematic heating is continuously present from spiral structure, bars, GMCs, etc. Therefore, they should be independent of the subtle difference in the numerical details of small-scale physics. A comprehensive analysis on disk morphology and its dependence on galaxy formation history using an enlarged sample of galaxies will be the subject of a future study. 

\begin{figure*}
\centering
\includegraphics[width=\linewidth]{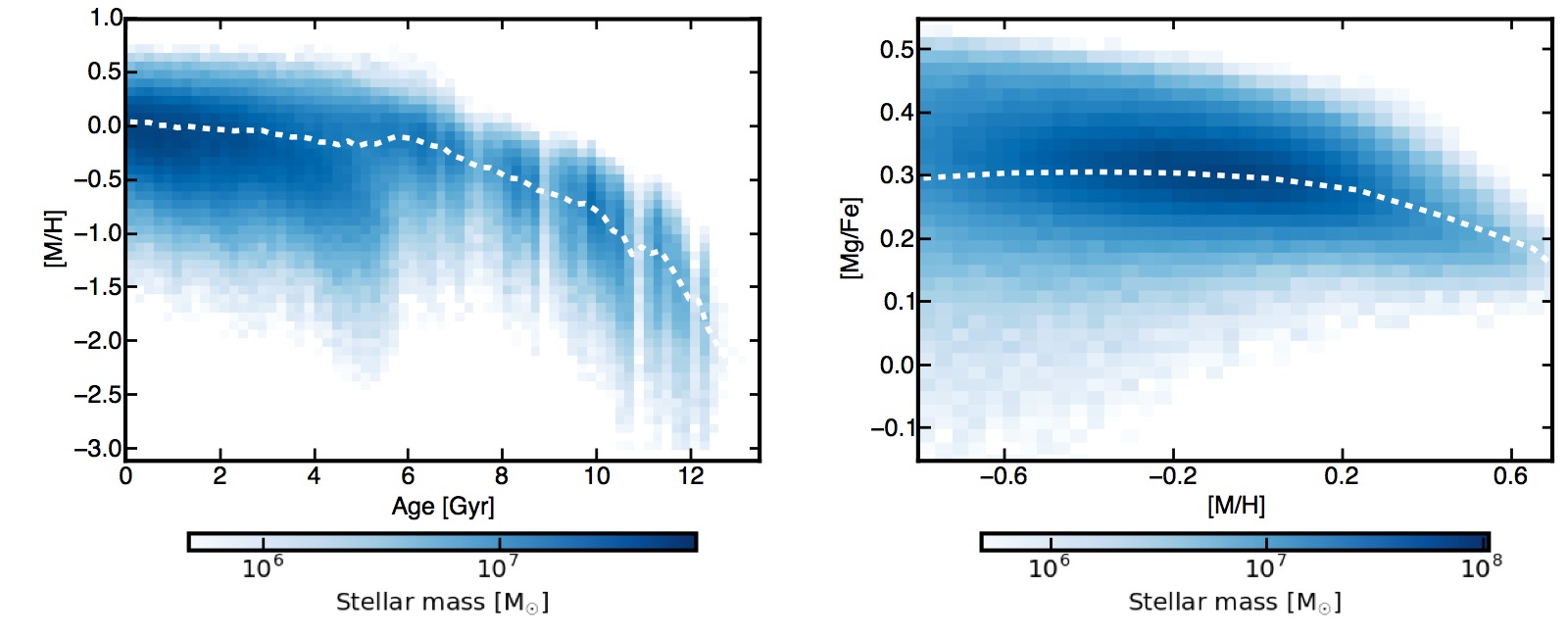}
\caption{{\em Left}: Stellar age--metallicity relation. {\em Right}: $\MgFe$--$\MH$ relation. Colors show the total stellar mass in each pixel (in logarithmic scale). The white dashed lines show the median for each relation. To leading order, stellar age, metallicity, and $\MgFe$ correlate with each other, but there is large scatter (over 1\,dex in $\MH$ at a given age).}
\label{fig:mono}
\end{figure*}

\subsection{The thin and thick disks}
\label{sec:thick}
In Section \ref{sec:disk}, we show that the vertical stellar density profile in the simulation can be well described by a two-component function (Fig. \ref{fig:profile}), which we refer to as the traditional thin disk ($\zh\sim200$--500\,pc) and thick disk ($\zh\sim1$--1.5\,kpc) \citep[e.g.][]{gilmore.reid.1983:thick.disk}. In terms of mass density, the thin and thick disks in our simulation can be roughly divided into stars younger and older than 4\,Gyr, respectively.

We first discuss the formation mechanisms of the thick disk. In Fig. \ref{fig:ZHAgeRad}, we show the scale heights as a function of stellar age at three radii $R=5$, 9, and 13\,kpc. The cross and plus symbols represent the scale heights of the thin and thick components, respectfully, as obtained from profile fitting. The vertical dotted line located at 4\,Gyr illustrates the separation of the thin and thick disks. The thick disk contains two distinct populations. First, about two thirds of the stars in the thick disk are older than 6\,Gyr (formation redshift $z\gtrsim0.7$). These were formed during the chaotic, bursty phase in the galaxy progenitor (Fig. \ref{fig:image}). This agrees with the picture proposed in \citet{brook.2004:thick.disk.form}. These stars have very large scale heights, as shown in Fig \ref{fig:ZHAgeRad}. Second, the other 1/3 of the stars in the thick disk are in the 4--6\,Gyr age interval, which were formed in a relatively calm, stable disk. The disk was more gas-rich and turbulent at early times, however, so the stars were formed thick `at birth', as proposed in \citet{bournaud.2009:thick.disk.form}. Furthermore, these stars continued to be kinematically heated into a thicker spatial distribution after forming. Therefore, the thick disk in our simulation is a mix of stars older than 4\,Gyr, which formed through a combination of several mechanisms.

Regarding the formation of the thin disk, we note that the gas disk smoothly became thinner down to $z=0$, thus forming the thin disk at late times. There is no sharp transition about 4\,Gyr ago when the thin disk at $z=0$ started to form. 

In the literature, some authors have claimed that there is a tension between preserving thin disks and the necessity of strong stellar feedback to prevent galaxies from forming too many stars in cosmological simulations \citep[e.g.][]{roskar.2014:effect.feedback.disk}. However, our simulation simultaneously forms a thin-disk component while producing a reasonable stellar mass and star formation history in good agreement with observational constraints \citep{hopkins.2014:fire.galaxy}. These results demonstrate that it is possible to form thin disks in cosmological simulations, even in the presence of strong stellar feedback. This is due to the fact that (1) our simulation has sufficient spatial resolution (smoothing length far less than the vertical scale heights of the thin disk), (2) we allow gas to cool to very low temperatures to explicitly resolve the cold, star-forming gas, and (3) the high resolution and the physically motivated star formation and feedback models adopted in the simulation allow one to explicitly resolve the launching and venting of galactic winds without disrupting the entire galaxy. Likewise, \citet{agertz.kravtsov.2015:feedback,agertz.kravtsov.2016:feedback} also found that their simulation can form a thin disk when using feedback recipes similar to ours but fail to do so using other feedback models.

\subsection{Stellar Migration in the Disk}
It has been proposed that radial migration of stars in the disk due to angular momentum exchange may be an important mechanism of disk heating that also flattens the stellar metallicity gradients \citep[e.g.][]{schonrich.binney.2009:migration,schonrich.binney.2009:structure,loebman.2011:thick.migration}. Recent numerical calculations suggest, however, that radial migration has little impact on the disk thickening \citep[e.g.][]{minchev.2012:migration.no.heat,martig.2014:disk.age.velocity,vera-ciro.2014:radial.migration,grand.2016:bar.heating,aumer.2016:disk.heat.gmc.spiral}. Nonetheless, radial migration can still occur when spiral arms and bars are present (via corotation resonance of transient spirals, e.g. \citealt{lynden-bell.kalnajs.1972:spiral,sellwood.binney.2002:migration}, or induced by long-lived spiral- or bar-like structures, e.g. \citealt{minchev.famaey.2010:resonance,minchev.2011:resonance.overlap}), while spiral arms and bars are suggested to be the dominant mechanism of disk heating \citep[e.g.][]{grand.2016:bar.heating}. 

In our simulation, stars older than 8\,Gyr show strong radial migration, because the bursty gas outflows driven by stellar feedback generate large fluctuations in the galactic potential, causing old stars to migrate toward large radius \citep{elbadry.2016:fire.migration}. This is important to shape the global structure of the disk \citep[e.g.][]{minchev.2015:thick.disk.formation}. However, this is a very different mechanism from the standard radial migration within a stable disk. To test the standard migration scenario, we study stars in the $z=0$ age interval 4--6\,Gyr. First, we go back to the snapshot at $\tb=4$\,Gyr and select stars in three annuli centered at $R=5$, 9, and 13\,kpc with 1\,kpc width near the disk plane ($\z<0.5$\,kpc) at that epoch. In the left panel in Fig. \ref{fig:migration}, we show the distribution of galactocentric radius $R$ for these stars by $z=0$. Stars in a given annulus 4\,Gyr ago have a wide distribution in $R$ by $z=0$. Only a small fraction (less than 10\%) of stars have migrated to very large radii ($\Delta R>5\,\kpc$), while more than half of the stars have migrated inward ($\Delta R<0$). This is expected from the exchange of angular momentum between stars, because outward-migrating stars carry more angular momentum, so more stars migrate inward. In the right panel, we show the stellar surface density and metallicity profiles in $R=4$--14\,kpc for all stars with $z=0$ age 4--6\,Gyr, measured at three epochs ($\tb=0,$ 2, and 4\,Gyr). The surface density does not change by more than 0.05\,dex during the past 4\,Gyr (stellar mass loss is subdominant). The average stellar metallicity has increased at large radii, resulting in a flattening of metallicity gradient by 0.01\,dex\,kpc$^{-1}$. Our results suggest that radial migration is common, but only has a weak net effect on the late-time global properties (mass density, metallicity profiles) of the galactic disk, consistent with predictions from other works \citep[e.g.][]{minchev.2013:chemodynamical,grand.kawata.2016:radial.migration}.

\begin{figure*}
\centering
\includegraphics[width=\linewidth]{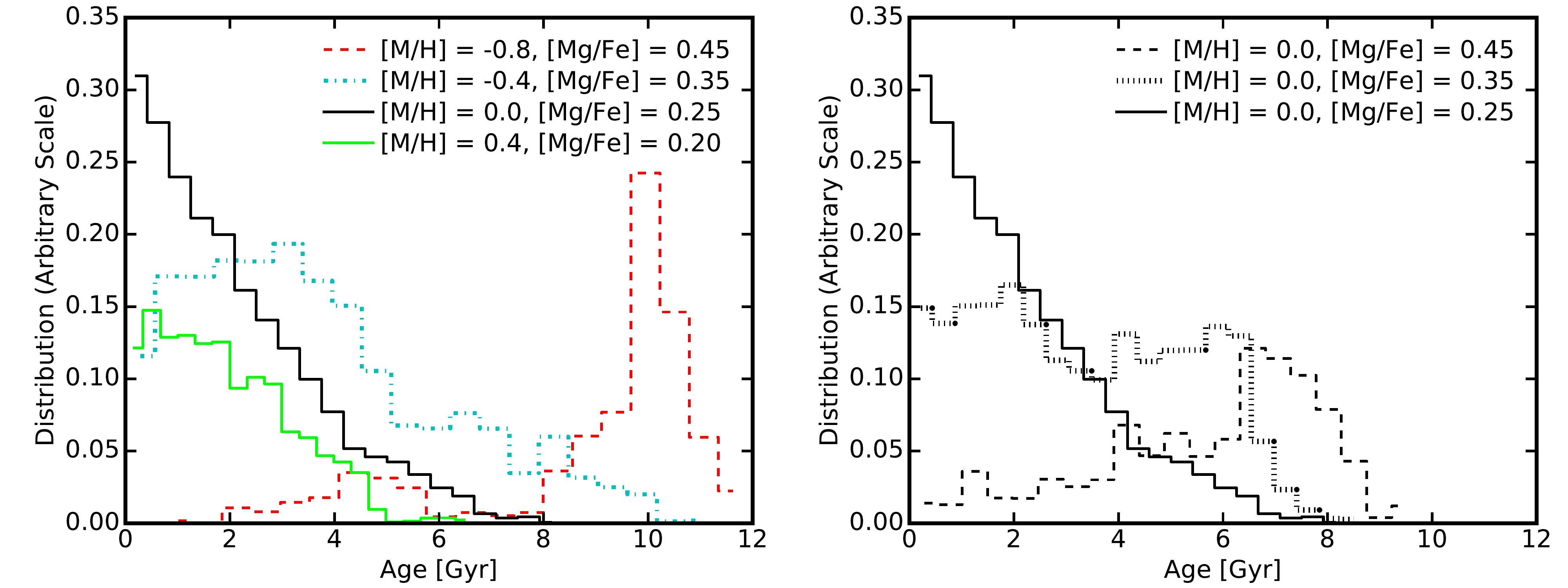}
\caption{Stellar age distribution of six mono-abundance populations in narrow bins with $\Delta\MH=0.06$ and $\Delta\MgFe=0.04$. In general, metal-poor, $\alpha$-rich stars represent older populations than metal-rich, $\alpha$-poor stars (left panel). At fixed metallicity, $\alpha$-poor stars are more biased toward younger populations (right panel). However, the age distribution of a mono-abundance population is wide.}
\label{fig:monoage}
\end{figure*}

\subsection{Abundance Patterns and Mono-abundance Populations}
\label{sec:mono}
In the literature, several authors have suggested that mono-abundance populations (stars with certain $\MH$ and $\aFe$) are proxies for single-age populations in the MW \citep[e.g.][]{bovy.2012:mono.structure,rix.bovy.2013:mw.disk.review}. This is important because one cannot infer the assembly history of the MW without reliable information on stellar ages. We first examine the abundance patterns in our simulation. In Fig. \ref{fig:mono}, we show the stellar age--metallicity relation in the left panel for all stars with $R=4$--14\,kpc and $\z<3\,\kpc$, color-coded by the total stellar mass in each pixel (in logarithmic scale). The white dashed line shows the median relation. In the right panel, we show the distributions of these stars in the $\MgFe$--$\MH$ plane. Stellar age, metallicity, and $\MgFe$ correlate with each other to leading order, but with considerable scatter (over 1\,dex in $\MH$ at a given age). The scatter mainly comes from the presence of metallicity gradients in the disk and non-uniform distribution of metals in the galaxy. We have explicitly checked that including sub-resolution metal diffusion in our simulation \citep[as in][]{shen.2010:metal.diffusion} does not dramatically reduce the scatter. 

In the left panel in Fig. \ref{fig:monoage}, we show the mass-weighted age distribution of stars from four mono-abundance populations selected within a tolerance in metallicity and abundance ratio of $\Delta\MH=0.06$ and $\Delta\MgFe=0.04$. In general, low-metallicity, $\alpha$-rich populations represent old stars, while high-metallicity, $\alpha$-poor stars are more biased toward younger populations. In the right panel, we compare three mono-abundance populations at fixed $\MH$ but different $\MgFe$. We find that low-$\alpha$ populations contain more young stars than $\alpha$-rich populations. These results marginally support the idea that chemical abundances might represent stellar ages to leading order. However, we caution that for any mono-abundance population, the age distribution is wide \citep[see also][]{minchev.2016:map.age.mw.disk}. For example, the most metal-poor and $\alpha$-rich population still contains a non-negligible fraction of young stars with age 2--6\,Gyr. As a consequence, if we repeat our analysis in this paper by breaking the stars into several bins of metallicity instead of stellar age, we obtain more complicated results due to age blending effects. Independent constraints on stellar age are required to break the degeneracy.

Recent observations reveal that MW stars fall into two distinct populations on the $\MgFe$--$\MH$ plane, known as the high- and low-$\alpha$ populations \citep[e.g.][]{adibekyan.2013:harps.fgk.dwarf,bensby.2014:fg.dwarf,anders.2014:chemo.apogee.dr1,nidever.2014:apogee.red.clump,mikolaitis.2014:ges.idr1.chemo,kordopatis.2015:ges.idr2.alpha}. Such feature is difficult to explain and cannot be reproduced by current cosmological simulations \citep[see][and reference therein]{nidever.2014:apogee.red.clump}, including ours. \citet{nidever.2014:apogee.red.clump} proposed several tentative models to explain how the two populations may form, but all require fine-tuned parameters to match the observed abundance pattern. Nevertheless, we show that in our simulation, star formation has undergone two distinct modes -- chaotic, bursty mode at high redshift and relatively calm, stable mode at late times. It is possible that such two-mode formation history may lead to a bimodality in the abundance pattern in very restricted conditions, while in other conditions the two populations may simply merge. Alternatively, the MW may have a very different assembly history from our simulation. In future work, we will study a large sample of simulations of MW-size halos with diverse assembly history and explore if and how such two populations form. Also, spectroscopic survey of stars in extragalactic galaxies with next generation of observational facilities may also reveal whether this is common in MW-mass galaxies or just a unique feature in the MW.

\begin{figure*}
\centering
\includegraphics[width=\linewidth]{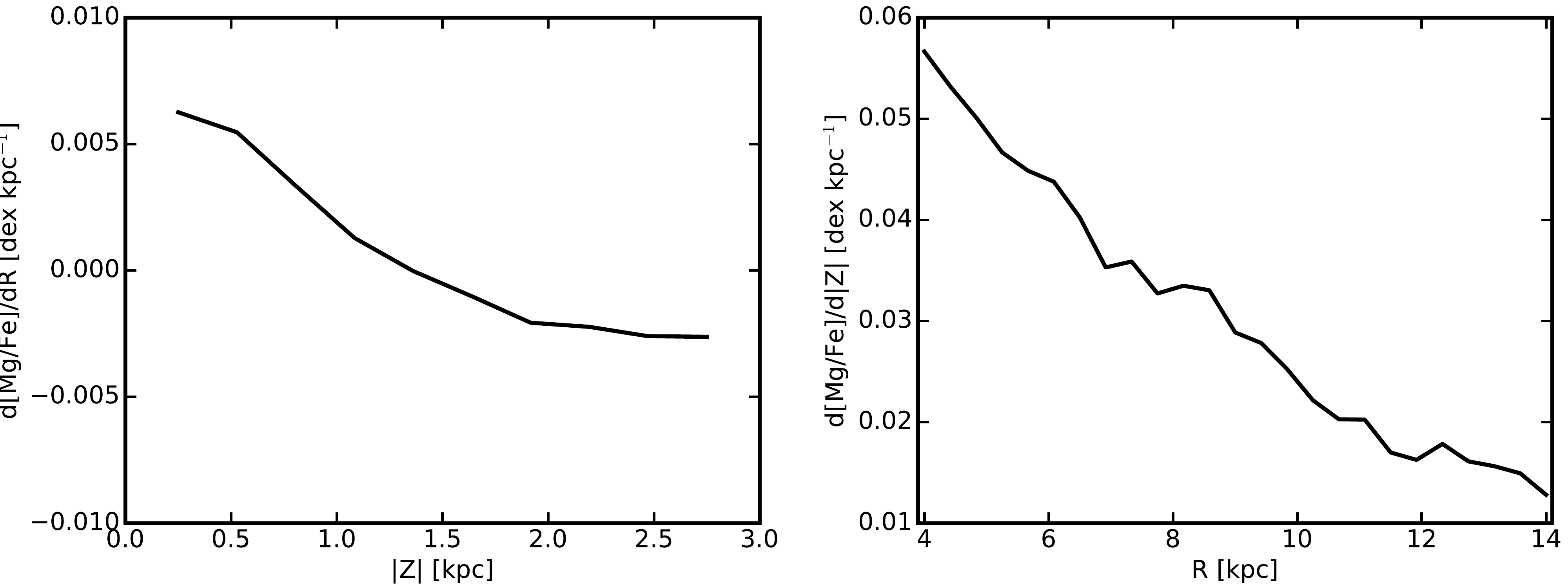}
\caption{{\em Left}: $\dd\MgFe/\dd R$ as a function of $\z$. {\em Right}: $\dd\MgFe/\dd\z$ as a function of $R$. The radial gradient in $\MgFe$ is positive at the mid-plane, flattens and becomes negative at larger heights. The vertical gradient in $\MgFe$ is positive, but is steeper at inner radii than at  outer radii. These trends are similar to those of metallicity gradient, but with flipped sign due to the anti-correlation between $\MgFe$ and metallicity. Both of them arise from the age gradient in the disk, in line with predictions from other simulations \citep[e.g.][]{minchev.2013:chemodynamical,minchev.2014:chemodynamical}.}
\label{fig:MgGrad}
\end{figure*}

\subsection{$\aFe$ gradients}
There is evidence indicating the presence of an $\aFe$ gradient in the MW disk \citep[e.g.][]{boeche.2014:rave.giant.grad,anders.2014:chemo.apogee.dr1,recio-blanco.2014:ges.thin.thick}, despite the fact that such measurements have large uncertainties. It is clear that the $\aFe$ gradient varies with $\z$ in the opposite way to $\FeH$ gradient and that $\dd\aFe/\dd R$ is negative at large $\z$, but the sign of $\dd\aFe/\dd R$ in the disk mid-plane are not fully consistent between various studies. Future spectroscopic surveys that include much larger samples will provide a more robust measurement. Here we use our simulation to make qualitative predictions for $\aFe$ gradients. In Fig. \ref{fig:MgGrad}, we show the radial gradient of $\MgFe$ as a function of height $\z$ (left panel) and the vertical gradient of $\MgFe$ as a function of radius $R$ (right panel). We find that $\dd\MgFe/\dd R$ is positive near the mid-plane, gradually decreases to zero at about $\z=1.3\,\kpc$, and turns negative at larger heights. Moreover, $\dd\MgFe/\dd\z$ is positive at all radii, but is larger at the inner disk than at the outer disk. These are qualitatively similar to the trends of metallicity gradient presented in Section \ref{sec:zgrad}, with the sign flipped following the anti-correlation between $\MgFe$ and $\MH$, as found also in other simulations \citep[e.g.][]{minchev.2013:chemodynamical,minchev.2014:chemodynamical}. The gradient of $\MgFe$ is also a consequence of the age gradient in the disk.

\section{Conclusions}
\label{sec:conclusion}
In this work, we have studied the structure, age and metallicity gradients, and dynamical evolution of the stellar disk via a case study of one simulation from the FIRE project, chosen to be a disk galaxy of similar mass to the MW at $z=0$. The simulation is a cosmological zoom-in simulation that includes physically motivated models of the multi-phase ISM, star formation, and stellar feedback, with parameters taken directly from stellar evolution models. Our main findings include the following:
\begin{enumerate}
\item Stars older than 6\,Gyr (formation redshift $z\gtrsim0.7$) were formed in a violent, bursty mode in a clumpy galaxy progenitor with powerful episodic outflows, and thus have round, puffy morphologies at $z=0$. Stars younger than 6\,Gyr were formed in a relatively calm, well-maintained star-forming disk. By $z=0$, stars that formed in the chaotic mode have the largest scale heights. Even for those that formed in a disk at late times, their scale heights increase with stellar age at any radius; stars of the same age have larger scale heights in the outer disk than in the inner disk (flaring). As a consequence, the median stellar age increases with $\z$ at a fixed radius, but decreases with $R$ in a constant-$\z$ layer.

\item The radial metallicity gradient is negative at the mid-plane, gradually flattens when moving to larger $\z$, and ultimately turns positive at about $\z>1.5$ kpc. The vertical metallicity gradient is negative at all radii, but is stronger at small radii. These trends are qualitatively consistent with observations in the MW. Such variation of metallicity gradient naturally follows the age gradient in the disk, since stellar age, metallicity, and $\aFe$ all correlate with each other. Similar trends also exist in $\aFe$ gradients, but with a flipped sign due to the anti-correlation between $\aFe$ and metallicity.

\item For stars that formed within the past 6\,Gyr, in a disk, those that formed earlier were thicker `at birth' than those formed later, because the star-forming disk was more gas-rich and therefore more turbulent and thicker at earlier times (a factor of $\sim2$ thicker at 6\,Gyr ago). After each population formed, their scale height was further increased via kinematic heating (by $\sim40\%$ during the past 6\,Gyr). In our simulation, the two factors have comparable effect in absolute units on the differential scale heights by $z=0$.

\item The vertical stellar density at $z=0$ can be well described by a two-component profile, defined as the traditional `thin disk' (scale heights $\zh\sim200$--500\,pc) and `thick disk' ($\zh\sim1$--1.5\,kpc). The thin and thick disks can be roughly separated by stars younger and older than 4\,Gyr. Two thirds of the stars in the thick disk are formed during the chaotic, bursty phase; the other 1/3 stars in the thick disk are formed in a gas-rich star-forming disk and then were further thickened via kinematic heating. The gas disk smoothly evolves during the past 6\,Gyr and forms the thin disk at late times. Therefore, the thick disk is a mix of stars that formed via different mechanisms, while the formation is continuous at the transition time (around 4\,Gyr ago) when the $z=0$ thin-disk stars started to form.

\item Our simulation demonstrates that it is possible to form a thin disk in sufficiently high-resolution cosmological simulations even in the presence of strong stellar feedback.
\end{enumerate}

Although we only study one simulation in this paper, our main results here are derived from global processes that can be understood with simple analytic arguments, including (1) star formation is bursty at high redshift and becomes relatively stable at late times, (2) the thickness of the star-forming gas disk decreases at low gas fraction, and (3) kinematic heating is continuously present from spiral structure, bars, GMCs, etc. In fact, some of our results agree very well with previous studies by other authors. For instance, almost all simulations of MW analogs show violent merger history at high redshift but relatively quiescent merger history at late times. Most of them form a separate thin and thick disk by $z=0$, with their scale heights and mass fractions similar to ours \citep[e.g.][]{brook.2012:thin.thick.disk.halo,bird.2013:disk.galaxy.assembly,minchev.2013:chemodynamical,martig.2014:disk.map.structure}. Some models also successfully reproduce the observed MW abundance distribution and the variation of stellar metallicity gradients, and attribute these results to the disk formation history \citep[e.g.][]{minchev.2013:chemodynamical,minchev.2014:chemodynamical}. Therefore, our results further confirm that the physical processes we proposed are common in the assembly histories of disk galaxies, regardless of numerical details and feedback model.

One key prediction of our simulation is that the thick disk does {\em not} form from a single channel, but it is rather a mixture of stars that formed and evolved in three different ways (see conclusion (iv) and Section \ref{sec:thick} for details). This scenario can be tested by future observations of MW stars if an independent constraint on stellar age can be obtained. For example, we expect the age separation between thin- and thick-disk stars is different from the one between stars that formed in the chaotic mode and in the calm mode. The latter can be identified by a sudden jump in the velocity dispersion as a function of stellar age.

Nevertheless, our simulation is not designed or chosen in any way to be identical to the MW and differs from it in several aspects. The radial metallicity gradient is $-0.03\,\dkpc$ in the disk mid-plane in our simulation, shallower than the $-0.06\,\dkpc$ slope in the MW disk \citep[e.g.][]{cheng.2012:segue.stellar.grad}. This can be affected by disk scale length, disk pre-enrichment at formation time, the extent of radial mixing, and specific merger or accretion events in the past. Also, our simulation does not show any bimodality in the $\aFe$--$\MH$ relation, in contrast to some observations of MW stars \citep[e.g.][]{nidever.2014:apogee.red.clump}. Moreover, our simulation does not show a prominent central bar, with which we expect that kinematic heating would be stronger \citep[e.g.][]{grand.2016:bar.heating}. Such differences may originate from details in the assembly history. In future work, we will further explore the disk formation, morphology, and metallicity profile and their dependence on galaxy formation history using an enlarged sample of disk galaxy simulations.

\section*{Acknowledgments}
We thank David Nidever, Hans-Walter Rix, Charlie Conroy, and Paul Torrey for useful discussions.
We also acknowledge Oscar Agertz and Ivan Minchev for helpful comments after the first draft has appeared on arXiv, and the anonymous referee for a detailed report.
The simulations used in this paper were run on XSEDE computational resources (allocations TG-AST120025, TG-AST130039, and TG-AST140023). 
The analysis was performed on the Caltech compute cluster ``Zwicky'' (NSF MRI award \#PHY-0960291).
Support for PFH was provided by an Alfred P. Sloan Research Fellowship, NASA ATP Grant NNX14AH35G, and NSF Collaborative Research Grant \#1411920 and CAREER grant \#1455342.
ARW was supported by a Caltech-Carnegie Fellowship, in part through the the Moore Center for Theoretical Cosmology and Physics at Caltech.
DAA acknowledges support by a CIERA Postdoctoral Fellowship.
CAFG was supported by NSF through grants AST-1412836 and AST-1517491, by NASA through grant NNX15AB22G, and by STScI through grants HST-AR-14293.001-A and HST-GO-14268.022-A.
DK was supported by NSF grant AST-1412153 and funds from the University of California, San Diego. 
EQ was supported by NASA ATP grant 12-APT12-0183, a Simons Investigator award from the Simons Foundation, and the David and Lucile Packard Foundation.

\bibliography{ms}

\appendix
\section{Resolution test}
\label{apd:res}
\begin{figure}
\centering
\includegraphics[width=\linewidth]{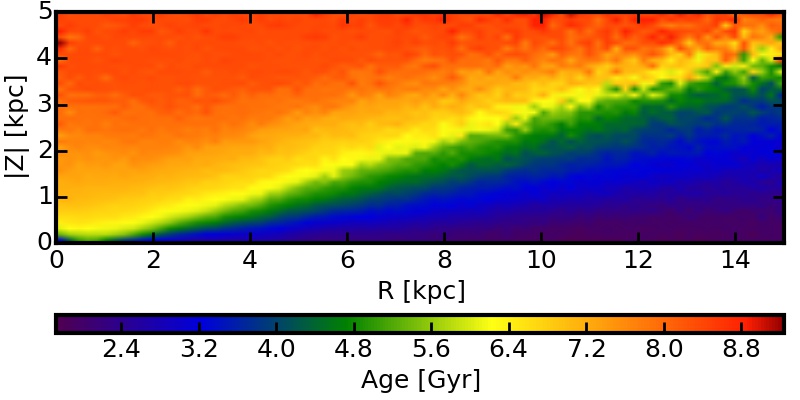} \\
\includegraphics[width=\linewidth]{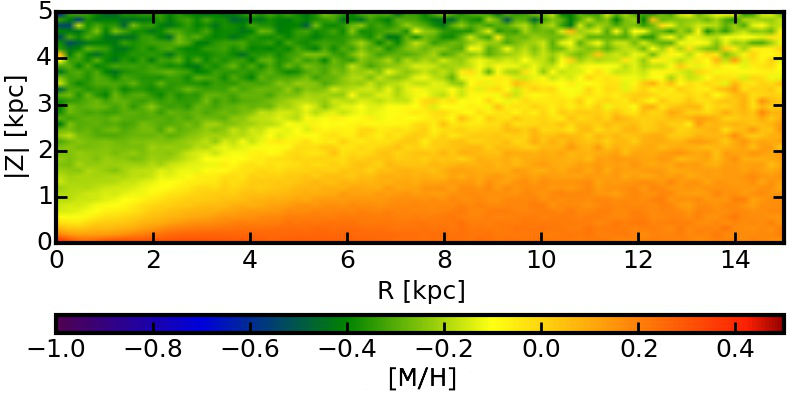} 
\caption{The same as Fig. \ref{fig:age}, but for the ultra-high-resolution simulation presented in \citet{wetzel.2016:high.res.mw.letter}. The disk structure does not significantly differ from the simulation studied in the paper, although this run has eight times better mass resolution and uses a more accurate hydrodynamic solver.}
\label{fig:ref13}
\end{figure}

In this paper, we performed a case study of a cosmological zoom-in simulation that produces a MW-mass disk galaxy at $z=0$. This simulation is originally presented in \citet{hopkins.2014:fire.galaxy} and has been thoroughly studied in other work \citep{vandevoort.2015:r.process.fire,faucher.2015:fire.neutral.hydrogen,muratov.2015:fire.mass.loading,ma.2017:fire.metallicity.gradient,ma.2016:fire.mzr,elbadry.2016:fire.migration}. Recently, \citet{wetzel.2016:high.res.mw.letter} have re-run this simulation with eight times better mass resolution and higher spatial resolution ($\epsilon_{\rm gas}=1\,\pc$ and $\epsilon_{\rm star}=4\,\pc$), but using the mesh-less finite-mass (MFM) hydrodynamics method in {\sc gizmo} and FIRE-2, an improved numerical implementation of the FIRE model (see Hopkins et al., in preparation, for details). We repeat our analysis on the new run and find all the results presented in the paper remain qualitatively unchanged. Particularly, the thin-to-thick disk decomposition, disk scale heights, and the amount of disk thickening at late times are consistent within 10\%. As one explicit example, in Fig. \ref{fig:ref13}, we show the median stellar age and average stellar metallicity as a function of $R$ and $\z$ in the new run. The disk structure and metallicity profile are very similar to the simulation analyzed in the paper (Fig. \ref{fig:age}). This suggests that our results are independent of resolution and numerical details, because most of the physics we consider in the paper are global processes and can be understood by simple analytic considerations. Nevertheless, there are some quantitative differences between the two runs. In the new run, the star-forming disk formed and stabilized at a later time ($\tb\sim4\,\Gyr$), due to stochastic effects during the last minor merger. The gas disk is more metal-enriched at formation time, so the radial metallicity gradient on the mid-plane is weaker. Moreover, the disk is more strongly flared, so the radial metallicity gradient turns positive at a lower height ($\z\sim1$\,kpc). This suggests that a large statistical sample is needed to make rigorous statements about quantitative details, as opposed to the robust qualitative trends we have focused on here.

\label{lastpage}

\end{document}